\documentclass[11pt]{article}

\usepackage{color}
\usepackage{amsmath, amsfonts,amssymb,theorem,euscript,
array,enumerate,amsfonts,mathrsfs}
\usepackage[T1]{fontenc}
\usepackage[latin1]{inputenc}
\usepackage{hhline}
\usepackage{graphicx}
\usepackage{array}
\usepackage{amsmath}
\usepackage{amssymb}
\usepackage{epstopdf}
\usepackage[latin1]{inputenc}
\usepackage[T1]{fontenc}
\usepackage{lmodern}
\usepackage{adjustbox}

\usepackage{eurosym}
\usepackage[font=small,skip=0pt]{caption}
\usepackage{dsfont}

\theoremheaderfont{\bfseries} \theoremstyle{plain}
\theorembodyfont{\upshape}

\newcommand{\nc}{\newcommand}
\nc{\ind}{\mathds{1}}

\def \trans{^{\scriptscriptstyle{\intercal}}}

\newcommand{\R}{\mathbb{R}}

\newcommand{\E}{\mathcal{E}}
\newcommand{\F}{\mathcal{F}}

\DeclareMathOperator{\esssup}{esssup}

\def\esssup_#1{\underset{#1}{\mathrm{ess\,sup\, }}}
\def\essinf_#1{\underset{#1}{\mathrm{ess\,inf\, }}}
\def\argmax_#1{\underset{#1}{\mathrm{arg\,max\, }}}
\def\argmin_#1{\underset{#1}{\mathrm{arg\,min\, }}}

\def \ep{\hbox{ }\hfill$\Box$}

\def \Prod{\displaystyle\prod}
\def \Int{\displaystyle\int}

\def\b1{\bf 1}

\def \R{\mathbb{R}}

\def \E{\mathbb{E}}
\def \F{\mathbb{F}}

\def \P{\mathbb{P}}

\def \S{\mathbb{S}}

\def \Ac{{\cal A}}

\def \Cc{{\cal C}}

\def \Ec{{\cal E}}
\def \Fc{{\cal F}}
\def \Gc{{\cal G}}

\def \Lc{{\cal L}}
\def \Pc{{\cal P}}

 \def \Rc{{\cal R}}
\def \Sc{{\cal S}}

\def \Vc{{\cal V}}
\def \Wc{{\cal W}}

\def \Vc{{\cal V}}

\def \eps{\varepsilon}

\def \ep{\hbox{ }\hfill$\Box$}

\newtheorem{Theorem}{Theorem}[part]

\newtheorem{Proposition}{Proposition}[part]

\newtheorem{Lemma}{Lemma}[part]

\newtheorem{Remark}{Remark}[part]

\def\reff#1{{\rm(\ref{#1})}}

\def\beqs{\begin{eqnarray*}}
\def\enqs{\end{eqnarray*}}
\def\beq{\begin{eqnarray}}
\def\enq{\end{eqnarray}}

\begin{document}

\title{Robust Markowitz mean-variance portfolio selection \\ under ambiguous 
covariance matrix
\thanks{This work is issued from a CIFRE collaboration between NATIXIS and LPMA.  We would like to thank Carmine De Franco, Johan Nicolle and Nizar Touzi  for helpful discussions. 
We are grateful to both referees and the AE for numerous comments, which help to improve the first version \cite{ismpha16} of this paper.}
}

\author{
Amine ISMAIL
\footnote{Natixis,  Equity Markets, and LPMA,Universit\'e Paris-Diderot,   \sf ami.ismael@gmail.com}
\qquad
Huy\^en PHAM%
\footnote{LPMA, Universit\'e Paris-Diderot and CREST-ENSAE, \sf  pham at math.univ-paris-diderot.fr.
The work of this author is part of the ANR project CAESARS (ANR-15-CE05-0024), and also supported by FiME 
and the ''Finance and Sustainable Development'' EDF - CACIB Chair.
}
}

\maketitle

\begin{abstract}
This paper studies a robust continuous-time Markowitz portfolio selection pro\-blem where the model uncertainty carries on  the 
covariance matrix of multiple risky assets.  
This problem is formulated into a min-max mean-variance problem over a set of non-dominated probability measures that is solved by a McKean-Vlasov dynamic programming approach, which allows us to characterize the solution in terms of a Bellman-Isaacs equation in the Wasserstein space of probability measures.  We provide explicit solutions for the optimal robust portfolio strategies and illustrate our results in the case of uncertain volatilities and ambiguous correlation between two risky assets. We then derive the robust efficient frontier in  closed-form, and obtain a lower bound for the Sharpe ratio of any robust efficient portfolio strategy. Finally, we  compare the performance of Sharpe ratios for a robust investor and for an investor with a misspecified model. 
\end{abstract}



\vspace{5mm}


\vspace{5mm}



\section{Introduction}

\setcounter{equation}{0}
\setcounter{Assumption}{0} \setcounter{Theorem}{0}
\setcounter{Proposition}{0} \setcounter{Corollary}{0}
\setcounter{Lemma}{0} \setcounter{Definition}{0}
\setcounter{Remark}{0}

The Markowitz mean-variance portfolio selection problem \cite{mar52}, initially considered in a single period model, is the cornerstone of modern portfolio allocation theory. Investment decisions rules are made according to the objective of maximizing the  expected return for a given financial risk quantified by the variance of the portfolio, and lead to the concept of  efficient frontier, which proposes a simple illustration of the trade-off  between  return and risk. 
The use of Markowitz efficient portfolio strategies in the financial industry has become quite popular mainly due to its natural and intuitive formulation. 

In a continuous-time dynamic setting, the mean-variance criterion involves in a nonlinear way the expected terminal wealth  due to the variance term, and induces the so-called time inconsistency. This nonstandard feature in stochastic control problem  has generated various resolution approaches. 
A first approach in \cite{zholi00} consists in embedding the mean-variance problem into an auxiliary standard  control problem that can be solved by using stochastic linear quadratic theory.  
A second approach relies on the observation that the dynamic mean-variance problem can be reformulated as a control problem of McKean-Vlasov type, where the cost functional may depend nonlinearly on  the law of the wealth state process.  
It has then been solved in \cite{anddje11} where the authors have derived a version of the Pontryagin maximum principle.  
More recently,  the paper \cite{phawei16} has developed a general dynamic programming approach for the control of McKean-Vlasov dynamics and applied their method for the  resolution of the mean-variance portfolio selection problem. 
We also mention the recent paper  \cite{fisliv15}, where the mean-variance problem is viewed as  the McKean-Vlasov limit of a family of controlled many-component weakly interacting systems. These prelimit problems are  solved by standard dynamic programming, and the solution to the original problem is obtained by passage to the limit. 

In the above cited papers,  the continuous-time Markowitz problem was essentially  stu\-died in the framework of a Black-Scholes model, and abundant research has been conducted to extend this setup by including models with random parameters. Among this large literature, we cite the recent paper 
\cite{chiwon14} which uses a stochastic correlation model for taking into account the correlation risk between risky assets. In all these works, it is assumed that investors have a perfect knowledge of the stochastic dynamics governing the price process, that is  a ``correct" model has to be first specified, and then the parameters have to be accurately estimated or calibrated. However, in finance, a model is clearly an appro\-ximation of the reality, and moreover within a model, the estimation problem is a difficult issue.  For example, it is known that the estimation of correlation between assets may be extremely inaccurate  due to asynchronous data and lead-lag effect, especially when the number of assets is large, and the correlation estimate converges to its true value less rapidly than the estimates of volatilities that are based on the full sets of marginal observations, see e.g. \cite{jagma03}, \cite{hayyos05} and  \cite{aitetal10}.   
On the other hand, optimal portfolios are typically  sensitive to the model and the parameters, and may perform badly when the parameters are not sufficiently accurate. Therefore, the impact of  model misspecification, due to erroneous models and measurements, 
is an important issue in the practical implementation of trading strategies, and is usually refereed to as model risk. 

In order to address the model risk related to uncertainty or ambiguous model parameters, the robust approach, which consists in  taking decisions under the worst-case scenario over all conceivable models, 
is a notable research direction in mathematical finance.  A common robust modeling is to consider a family of probability measures representing all the prior beliefs of the investor on the model parameters.  For example, drift uncertainty is modeled via Girsanov's theorem by a set of dominated probability measures, and has been first considered in the context of portfolio selection in \cite{hansar01}, and then largely studied in the literature,  see the recent paper  \cite{jinzho15} and  the references therein.

We focus here on uncertainty or ambiguity on the covariance matrix of the risky assets, assuming that the instantaneous return (drift)  
is known (or by considering that we have a strong belief on its value).  
Uncertain volatility models have been considered in 
\cite{aveetal95}, \cite{lyo95}, or \cite{denmar06} in the context of option pricing, and in \cite{maetal15}, \cite{linrie14} for robust portfolio optimization with expected utility criterion.  
As in \cite{fouetal15}, we are also interested in a setting with ambiguous  correlation between two risky assets since, as already mentioned above, 
the correlation parameter is hard in practice to infer with accuracy from market information.  

In this paper, we investigate the  robust Markowitz mean-variance portfolio selection under uncertainty on the volatilities and correlation of multi  risky assets. Robust mean-variance problems have been considered in the economic and engineering literature, mostly on single period or multiperiod models, see e.g. \cite{fabetal10}, \cite{pin16} and \cite{liuzhe16}.
Here, in our continuous-time modeling, we adopt the probabilistic framework in \cite{epsji13}, related to the theory of $G$-expectation \cite{pen06} (see also \cite{sonetal12}), in order to capture model uncertainty and ambiguity on the covariance matrix,  which leads to a set of non-dominated probability measures for the prior probabilities.  We also make some concavity assumption on the set of prior covariance matrix. 
From a mathematical viewpoint, and compared to robust problem with expected utility, we face two additional difficulties: (i) it cannot be tackled a priori by classical stochastic differential game  approach due to the nonlinear variance term, (ii)  moreover, since the worst-case scenario is not the same for the mean and the variance, it is not straightforward that it can be put into a min-max problem.  We then use  the following methodology. 
We consider a robust mean-variance criterion, which is actually formulated as a min-max problem, and show a posteriori how it is connected to the robust Markowitz problem. We tackle the former problem by a McKean-Vlasov dynamic programming approach: we first reformulate the robust mean-variance problem into a deterministic differential game problem with the law of the wealth process under a prior probability measure as state variable. 
Then, adapting optimality arguments  from dynamic programming principle, and using recent chain rule for flow of probability measures 
derived in \cite{buetal14} and \cite{chacridel15}, we state a verification theorem which gives the optimal strategy and  performance in terms of a Bellman-Isaacs equation in the Wasserstein space of probability measures. We next apply this analytic partial differential equation characterization of the solution to the robust mean-variance problem, and show that the problem  
can be reduced into two steps: first, we determine the worst-case scenario, and the remarkable point is that it corresponds to a constant variance/covariance matrix  
obtained by the minimization of the risk premium, which is a direct input of the model.   
Secondly, we obtain the optimal mean-variance strategy as in the Black-Scholes model with the known instantaneous return and the worst-case constant covariance matrix. We 
illustrate our results with closed-form expressions for the optimal portfolio strategies  in two examples: uncertain volatilities and ambiguous correlation between two risky assets. 
Moreover, we are  able to derive explicitly the corresponding robust efficient frontier of the robust Markowitz problem.  In particular, we obtain a lower bound for the Sharpe ratio of any robust efficient portfolio strategy, which is independent of any modelling on the covariance matrix. 

How   can  robust mean-variance portfolio strategies help to improve performance of investors? 
We address this question by  using simulations to evaluate and compare  the Sharpe ratio of a robust investor and a  simple investor who implements mean-variance strategies with a misspecified model in two examples: (i) in the first example, the true dynamics of the stock price is assumed to be governed by a Heston type stochastic volatility model that makes the volatility bounded, and the simple investor considers that the risky asset is governed by a Black-Scholes model  with constant volatility,  (ii) in the second example, the two-assets price is given in reality by a stochastic correlation model, but the simple investor considers a constant correlation between the risky assets. Our results show that the robust Sharpe ratio can perform noticeably better than the misspecified Sharpe ratio  
for some  choice of the parameters describing the true dynamics.

The rest of the paper is organized as follows. Section 2 formulates the probabilistic framework for the robust Markowitz mean-variance problem. 
We present in Section 3 the McKean-Vlasov dynamic programming approach for solving our problem. In Section 4, we derive explicit solutions in the context of ambiguous covariance matrix including uncertain volatilities and ambiguous correlation. Section 5 is devoted to the derivation of the robust efficient frontier in closed form, and the last Section 6 discusses the benefit of a robust investor compared to a misspecified investor.

\section{Problem formulation} \label{secprob}

\setcounter{equation}{0}
\setcounter{Assumption}{0} \setcounter{Theorem}{0}
\setcounter{Proposition}{0} \setcounter{Corollary}{0}
\setcounter{Lemma}{0} \setcounter{Definition}{0}
\setcounter{Remark}{0}

We consider a financial market with one risk-free asset, assumed to be constant equal to one (zero interest rate),  and $d$ risky stocks on a finite investment horizon $[0,T]$. 
We model the uncertainty about the volatility matrix of the risky assets by using  the probabilistic setup as in \cite{denmar06}, \cite{pen06} or \cite{sonetal12}. We define the canonical state space by  $\Omega$ $=$ $\{ \omega = (\omega(t))_{t\in [0,T]} \in C([0,T];\R^n): \omega(0) = 0\}$ representing the continuous paths driving  $d$ risky assets, and possibly $m$ (non tradable) factor processes ($n$ $=$ $d+m$), by $\Fc$ its Borel 
$\sigma$-field,  and denote by $\bar B$ $=$ $(\bar B_t)_{t\in [0,T]}$ the canonical process, i.e. $\bar B_t(\omega)$ $=$ $\omega(t)$, 
by  $\P_0$  the Wiener measure, i.e. making $\bar B$ a $n$-dimensional Brownian motion under $\P_0$, and by $\F$ $=$ $(\Fc_t)_{0\leq t\leq T}$ the canonical filtration, i.e. the natural filtration generated by $\bar B$. 
We distinguish the $d$-dimensional components of $\bar B$, denoted by $B$, and 
representing the continuous paths of the risky assets, and the other $(n-d)$-dimensional components are denoted by $\check B$.

The investor knows  (or has estimated) the constant drift $b$ $=$ $(b_1,\ldots,b_d)$ $\in$ $\R^d$ of the assets, but  is uncertain about the  volatility matrix  (possibly random) of the $d$ risky assets.  
We adopt the concept of ambiguous  volatility as defined in \cite{epsji13}, 
which means that the investor only knows that the covariance matrix  belongs to some prior  compact set  $\Gamma$ of  $\S_{>+}^{d}$,  
the set of strictly positive definite matrices in $\R^{d\times d}$. We assume  that  $\Gamma$ $=$ $\Gamma(\Theta)$  
is parametrized by a prior convex set $\Theta$ of $\R^q$, that is there exists some measurable function $\gamma$ $:$ 
$\R^q$ $\rightarrow$ $\S_{>+}^{d}$ s.t. any $\Sigma$ $\in$ $\Gamma$ is in the form $\Sigma$ $=$ $\gamma(\theta)$ for some $\theta$ $\in$ $\Theta$.  
For any $\Sigma$ $\in$ $\Gamma$, we denote by $\sigma$ $=$ $\Sigma^{1\over 2}$ its square-root matrix, and we shall often identify a covariance matrix with its square-root matrix called volatility matrix.   Here are some examples of this modeling:

\vspace{2mm}

\noindent {\bf Example 1 (uncertain volatilities)}.   In dimension $d$ $=$ $1$, this is modelled through $\Gamma$ $=$ $\Theta$ $=$ 
$[\underline{\sigma}^2,\bar\sigma^2]$ with positive constants $0$ $<$ $\underline{\sigma}$ $\leq$ $\bar\sigma$ $<$ $\infty$, see \cite{aveetal95}, \cite{lyo95}.  The extension to the multivariate assets case with zero correlation is modelled through 
$\Theta$ $=$ $\Prod_{i=1}^d[\underline{\sigma_i^2},\bar\sigma_i^2]$ with 
$0<\underline{\sigma_i}$ $\leq$ $\bar\sigma_i$ $<$ $\infty$, $i$ $=$ $1,\ldots,d$, and 
\beqs
\gamma(\theta) &=& 
\left(
\begin{array}{ccc}
\sigma_1^2 & \ldots & 0 \\
\vdots & \ddots & \vdots \\
0 & \ldots & \sigma_d^2
\end{array}
\right), \;\;\; \mbox{ for } \; \theta = (\sigma_1^2,\ldots,\sigma_d^2). 
\enqs

\vspace{2mm}

\noindent {\bf Example 2 (ambiguous correlation)}. The uncertainty about the correlation between risky assets in dimension $d$ $=$ $2$ has been recently considered in \cite{fouetal15}, and can be formalized here 
with $\Theta$ $=$ $[\underline{\varrho},\bar\varrho]$ $\subset$ $(-1,1)$, and 
\beqs
\gamma(\theta) \; = \;  \left( 
\begin{array}{cc}
\sigma_1^2   & \sigma_1 \sigma_2 \theta \\
 \sigma_1 \sigma_2 \theta   & \sigma_2^2 
\end{array}
\right),
\enqs
for some known positive constants $\sigma_1$ and $\sigma_2$ representing the marginal volatilities of the assets, and where $\theta$  represents the unknown correlation parameter varying between $\underline{\varrho}$ and $\bar\varrho$. 
The extension to multivariate assets for $d$ $\geq$ $2$ can also be done within our framework with a parametric form for the correlation matrix using for instance $d(d-1)/2$  angular coordinates as in \cite{rebjac99}.

\vspace{2mm}

We denote by $\Vc_\Theta$ the set of $\F$-progressively measurable processes $\Sigma$ $=$ $(\Sigma_t)$ valued in $\Gamma$ $=$ 
$\Gamma(\Theta)$, and introduce the set of prior probability measures $\Pc^\Theta$: 
\beqs
\Pc^\Theta &=& \big\{ \P^\sigma: \Sigma \in \Vc_\Theta \big\}, 
\enqs
where $\P^\sigma$ is the probability measure on $(\Omega,\Fc_T)$ induced by $\P_0$ via:
\beqs
\P^\sigma &:=& \P_0  \circ (\bar B^\sigma)^{-1},\;\;\;  \mbox{ with } \sigma_t := \Sigma_t^{1\over 2}, 
\; B_t^\sigma :=  \int_0^t \sigma_s dB_s, \;\; 0 \leq t \leq T, \; \P_0-a.s.,
\enqs
and $\bar B^\sigma$ is the $\R^n$-valued process on $\Omega$ defined by  $\bar B^\sigma$ $:=$ $(B^\sigma \; \check B)$. 

Under any $\P^\sigma$, $\Sigma$ $\in$ $\Vc_\Theta$, the process $B$ is a martingale, hence admits from \cite{kar95}, a  quadratic variation, which is given by:  
\beqs
d<B>_t &=& \Sigma_t dt.
\enqs

\begin{Remark}
{\rm Ambiguity in volatility leads to a set of prior probabilities in $\Pc^\Theta$, which are  non-equivalent, actually  mutually singular.  Such a specification for the set of prior probabilities $\P^\sigma$  
is closely connected to the theory of $G$-Brownian motion introduced in \cite{pen06}, and requires  tools from quasi-sure analysis as pointed out in \cite{denmar06}, and further studied in \cite{sonetal12}. 
In particular, we say that a property holds $\Pc^\Theta$-quasi surely ($\Pc^\Theta-q.s$. in short), if it holds  $\P^\sigma-a.s.$  for all $\P^\sigma$ $\in$ $\Pc^\Theta$.
}
\ep
\end{Remark}

\vspace{2mm}

The price process $S$ of the $d$ risky assets is given by 
\beqs
dS_t &=& {\rm diag}(S_t) \big( b dt + dB_t), \;\;\; 0 \leq t \leq T, \;\; \Pc^\Theta-q.s. 
\enqs

\begin{Remark} \label{remS} 
{\rm Under each $\P^\sigma$  $\in$ $\Pc^\Theta$, for $\Sigma$ $\in$ $\Vc_\Theta$, we have $dB_t$ $=$ $\sigma_t dW_t^\sigma$ where $W^\sigma$ is a   Brownian motion under $\P^\sigma$, and so the price process is governed under 
$\P^\sigma$ by
\beqs
dS_t &=& {\rm diag}(S_t) \big( b dt + \sigma_t dW_t^\sigma), \;\;\; 0 \leq t \leq T, \;\; \P^\sigma-a.s. 
\enqs
}
\ep
\end{Remark}

A portfolio  strategy  $\alpha$ $=$ $(\alpha_t)_{0\leq t\leq T}$, representing the amount invested in the $d$ risky assets, is a $d$-dimensional  $\F$-progressively measurable  process, valued in some closed convex set $A$ of $\R^d$, satisfying the integrability condition
\beq \label{integalpha}
\sup_{\P^\sigma\in\Pc^\Theta} \E_{\sigma} \Big[ \int_0^T \alpha_t\trans\Sigma_t\alpha_t dt \Big] & < & \infty, 
\enq
and denoted by $\alpha$ $\in$ $\Ac$.  Here $\trans$ denotes the transpose of a matrix, and $\E_\sigma$ denotes the expectation under $\P^\sigma$.  Given a portfolio strategy $\alpha$ $\in$ $\Ac$, and an initial capital $x_0$ $\in$ $\R$, 
the evolution of the self-financing wealth process $X^\alpha$ is given by
\beq \label{Xagreg}
dX_t^\alpha &=&\alpha_t\trans {\rm diag}(S_t)^{-1} dS_t \; = \;  \alpha_t\trans\big( bdt + dB_t), \;\;\ 0\leq t\leq T, \; X_0^\alpha = x_0, \; \Pc^\Theta-q.s. 
\enq

\begin{Remark}
{\rm Given $\alpha$ $\in$ $\Ac$, the existence of a $\Pc^\Theta$-quasi surely aggregated solution to \reff{Xagreg} is ensured by Theorem 2.2 in \cite{nut12} 
under the Zermelo Fraenkel set theory with axiom of choice (ZFC) plus the Continuum Hypothesis.
Moreover, for $\alpha$ $\in$ $\Ac$, and from Remark \ref{remS}, we see that the evolution of $X^\alpha$ under any $\P^\sigma$ $\in$ $\Pc^\Theta$,  
$\Sigma$ $\in$ $\Vc_\theta$, is given by 
\beq \label{dynXsig}
dX_t^{\alpha} &=& \alpha_t\trans (b dt + \sigma_t dW_t^\sigma), \;\;\; 0 \leq t\leq T, \; X_0^{\alpha}  = x_0, \; \P^\sigma-a.s. 
\enq
where $W^\sigma$ is a Brownian motion under $\P^\sigma$, and we have
\beqs
\sup_{\P^\sigma\in\Pc^\Theta} \E_{\sigma} \big[ \sup_{0\leq t\leq T} |X_t^\alpha|^2 \big] & < & \infty. 
\enqs 
}
\ep
\end{Remark}

\vspace{2mm}

Given a risk aversion parameter $\lambda$ $>$ $0$, the  worst-case mean-variance functional under ambiguous volatility  is 
\beqs
J_{wc}(\alpha) &=& \sup_{\P^\sigma\in\Pc^\Theta} \Big( \lambda {\rm Var_\sigma}(X_T^\alpha) - \E_\sigma[X_T^\alpha] \Big) \; < \; \infty,  \;\;\; \alpha \in \Ac, 
\enqs
where ${\rm Var}_\sigma(X)$ denotes the variance of $X$ under $\P^\sigma$, and the robust mean-variance portfolio selection problem   is then formulated as 
\beq \label{robustMV}
V_0 & = &  \inf_{\alpha\in\Ac} J_{wc}(\alpha) \; = \; \inf_{\alpha\in\Ac} \sup_{\P^\sigma\in\Pc^\Theta} \Big( \lambda {\rm Var_\sigma}(X_T^\alpha) - \E_\sigma[X_T^\alpha] \Big). 
\enq

A related problem to the robust mean-variance portfolio selection problem is the robust Markowitz problem, which is formulated as follows:  
given a variance risk $\vartheta$ $>$ $0$, 
\begin{equation} \label{robustMarkowitz}
\left\{
\begin{array}{rcl}
\mbox{maximize  over } \alpha \in \Ac, & & \Ec(\alpha)  \; := \; \inf_{\P^\sigma\in\Pc^\Theta}  \E_\sigma[X_T^\alpha] \\
\mbox{subject to} & &  \Rc(\alpha) \; := \; \sup_{\P^\sigma\in\Pc^\Theta} {\rm Var_\sigma}(X_T^\alpha) \; \leq  \; \vartheta.  
\end{array}
\right.
\end{equation}
A solution $\hat\alpha^\vartheta$ to \reff{robustMarkowitz}, when it exists, is called {\it robust efficient portfolio strategy} with respect to $\vartheta$. 
In other words, a robust efficient portfolio strategy maximizes the  worst case expected terminal wealth given a financial risk measured by  the worst case variance of the terminal wealth.  The pair  $(\Rc(\hat\alpha^\vartheta),\Ec(\hat\alpha^\vartheta))$  is called a robust efficient point, and the 
set of all robust efficient points, when varying $\vartheta$,  is called {\it robust efficient frontier}. 
By standard convex optimization theory, the constrained optimization problem \reff{robustMarkowitz} is connected by duality to the Lagrangian optimization problem, which is defined as
\beqs
\inf_{\alpha\in\Ac}  \big[ \lambda \Rc(\alpha) - \Ec(\alpha) \big]  &=& \inf_{\alpha\in\Ac} \Big\{ \lambda  \sup_{\P^\sigma\in\Pc^\Theta}  {\rm Var_\sigma}(X_T^\alpha) -  
\inf_{\P^\sigma\in\Pc^\Theta}  \E_\sigma[X_T^\alpha]  \Big\}.
\enqs
Notice that this Lagrangian optimization problem is equal to problem \reff{robustMV} when $\Pc^\Theta$ is  a singleton, but differs a priori  from \reff{robustMV}.  We shall solve in the two next sections the robust mean-variance portfolio selection problem \reff{robustMV}, and show in the last section that it is actually equal by duality to the Lagrangian optimization problem, and so  leads to the solution of the robust Markowitz problem \reff{robustMarkowitz} and the construction of the robust efficient frontier.

\section{McKean-Vlasov approach}

\setcounter{equation}{0}
\setcounter{Assumption}{0} \setcounter{Theorem}{0}
\setcounter{Proposition}{0} \setcounter{Corollary}{0}
\setcounter{Lemma}{0} \setcounter{Definition}{0}
\setcounter{Remark}{0}

 Problem \reff{robustMV} can be viewed as a zero-sum  stochastic differential game problem with  gain/cost functional 
\beq \label{JMV}
J(\alpha,\sigma) &=& \lambda {\rm Var_\sigma}(X_T^\alpha) - \E_\sigma[X_T^\alpha], \;\;\: \alpha \in \Ac, \Sigma \in \Vc_\Theta, 
\enq
so that $V_0$ $=$ $\inf_{\alpha\in\Ac} \sup_{\Sigma\in\Vc_\Theta} J(\alpha,\sigma)$.  The peculiarity of this differential game problem is the nonlinear dependence of the law of the state process via the variance term, 
making the problem a priori time inconsistent. Following the idea in \cite{benetal15} and \cite{phawei16} for control problem, we first reformulate our problem into a deterministic differential game problem, taking into account the uncertainty about the probability law governing the risky asset. 
For  any $\alpha$ $\in$ $\Ac$, and $t$ $\in$ $[0,T]$, let us denote by $\rho_t^{\alpha,\sigma}$ $=$ $\P_{_{X_t^\alpha}}^\sigma$ the law of $X_t^\alpha$ under $\P^\sigma$, $\Sigma \in \Vc_\Theta$, 
which defines  a deterministic process valued in the Wasserstein space $\Pc_{_2}(\R)$ of square-integrable probability measures on $\R$, 
which is a metric space when equipped with the Wasserstein distance $\Wc_{_2}$: 
\beqs
\Wc_{_2}(\mu,\mu') &=& \inf\Big\{ \Big( \int_{\R\times \R} |x-y|^2 \pi(dx,dy)\Big)^{1\over 2}:
\pi \in \Pc_{_2}(\R\times \R) \mbox{ with marginals } \mu \mbox{ and } \mu' \Big\} 
\enqs
We also set $\|\mu\|_{_2}$ $:=$ $\Wc_{_2}(\mu,\delta_0)$ $=$ $\big(\int |x|^2 \mu(dx)\big)^{1\over 2}$.

We also introduce the following convenient notations: for any $\mu$ $\in$ $\Pc_{_2}(\R)$, we denote by 
\beqs
\bar\mu \; := \;  \int_\R x \mu(dx), & &  {\rm Var}(\mu) \; := \; \int_{\R} (x - \bar\mu)^2 \mu(dx). 
\enqs
We can then rewrite the functional in \reff{JMV} and the worst-case mean-variance functional  as 
\beq \label{Jdet}
J_{wc}(\alpha) \; = \; \sup_{\Sigma\in\Vc_\Theta} J(\alpha,\sigma) &=&  
\sup_{\Sigma\in\Vc_\Theta} \big[ \lambda {\rm Var}(\rho_T^{\alpha,\sigma}) - \overline{\rho_T^{\alpha,\sigma}}\big], \;\;\; \alpha\in\Ac. 
\enq
The robust mean-variance portfolio selection problem  is therefore reformulated as a  deterministic differential game problem with controlled state variable $\rho^{\alpha,\sigma}$ valued in the infinite-dimensional space 
$\Pc_{_2}(\R)$. To solve this problem, we use  general dynamic programming optimality principle, which takes the following formulation in our context:

\vspace{2mm}
 
\noindent {\bf Optimality principle} 

\vspace{1mm}

\noindent  Let $\{V^{\alpha,\sigma},\alpha\in\Ac,\Sigma\in\Vc_\Theta\}$ be a family  of deterministic processes in the form $V_t^{\alpha,\sigma}$ $=$ $v(t,\rho_t^{\alpha,\sigma})$ for some real-valued measurable function 
$v$ on $[0,T]\times\Pc_{_2}(\R)$ satisfying
\begin{itemize}
\item[(i)]  $v(T,\mu)$ $=$ $\lambda {\rm Var}(\mu) - \bar\mu$, for any $\mu$ $\in$ $\Pc_{_2}(\R)$
\item[(ii)] $t$ $\in$ $[0,T]$ $\longmapsto$ $\sup_{\Sigma\in\Vc_\Theta} V_t^{\alpha,\sigma}$ is nondecreasing for all $\alpha$ $\in$ $\Ac$
\item[(iii)] $t$ $\in$ $[0,T]$ $\longmapsto$ $\sup_{\Sigma\in\Vc_\Theta} V_t^{\alpha^*,\sigma}$ is nonincreasing (hence constant) for some 
$\alpha^*$ $\in$ $\Ac$.
\end{itemize}
Then, $\alpha^*$ is an optimal control for the robust mean-variance problem \reff{robustMV} with optimal value
\beq \label{optalpha*}
V_0  \; = \; v(0,\delta_{_{x_0}}) &=& J_{wc}(\alpha^*).  
\enq

Indeed,  observe that  at time $t$ $=$ $0$, $\rho_0^{\alpha,\sigma}$ $=$ $\delta_{_{x_0}}$ for any $\alpha \in \Ac, \Sigma \in \Vc_\Theta$, since 
$X_0^\alpha$ is equal to the constant $x_0$, which implies that 
$V_0^{\alpha,\sigma}$ $=$  $v(0,\delta_{_{x_0}})$ does not depend on $\alpha \in \Ac, \Sigma \in \Vc_\Theta$.  From properties (i) and (ii), we then have for all $\alpha$ $\in$ $\Ac$, 
\beqs
v(0,\delta_{_{x_0}})  \; = \; \sup_{\Sigma\in\Vc_\Theta} V_0^{\alpha,\sigma} 
& \leq & \sup_{\Sigma\in\Vc_\Theta} V_T^{\alpha,\sigma} \; = \;  \sup_{\Sigma\in\Vc_\Theta} v(T,\rho_T^{\alpha,\sigma}) \; = \; 
\sup_{\Sigma\in\Vc_\Theta} J(\alpha,\sigma) \; = \;  J_{wc} (\alpha),
\enqs
by \reff{Jdet}. Since $\alpha$ is arbitrary in $\Ac$, this gives:  $v(0,\delta_{_{x_0}})$ $\leq$ $\inf_{\alpha\in\Ac}  J_{wc}(\alpha)$ $=$ $V_0$. 
Similarly, from properties (i) and (iii), we obtain $v(0,\delta_{_{x_0}})$  $=$  $\sup_{\Sigma\in\Vc_\Theta}  J(\alpha^*,\sigma)$ $=$ $J_{wc}(\alpha^*)$ 
$\geq$ $V_0$, 
which proves \reff{optalpha*}.

\vspace{3mm}

In order to construct a process $V_t^{\alpha,\sigma}$ $=$ $v(t,\rho_t^{\alpha,\sigma})$ satisfying the above  conditions (i), (ii), (iii) for the optimality principle, we shall rely on the recent notion of derivatives in the Wasserstein space introduced  by P.L. Lions, and the corresponding chain rule (It\^o's formula) for flow of probability measures, that we recall in the appendix.  The derivative (when it exists) of a function $\varphi(\mu)$ on $\Pc_{_2}(\R)$ is denoted by $\partial_\mu\varphi(\mu)$, and is a function 
from $\R$ into $\R$, which is in $L^2(\mu)$,  and when a version of the function $x$ $\in$ $\R$  $\mapsto$ $\partial_\mu\varphi(\mu)(x)$ is differentiable, we denote by $\partial_x\partial_\mu\varphi(\mu)(x)$ its derivative. 
Assuming that $v(t,\mu)$ is smooth on $[0,T]\times\Pc_{_2}(\R)$, i.e. continuously differentiable w.r.t. to  $t$, and partially $\Cc^2$ w.r.t. $\mu$ (see Appendix B), 
we have by It\^o's formula \reff{Ito} (recalling \reff{dynXsig}): 
\beq \label{dV}
dV_t^{\alpha,\sigma} \; = \; d v(t,\rho_t^{\alpha,\sigma}) &=&  D_t^{\alpha,\sigma} dt, 
\enq
where 
\beq \label{defD}
D_t^{\alpha,\sigma} &=& \partial_t v(t,\rho_t^{\alpha,\sigma})  + \E_\sigma \big[  H(\partial_\mu v(t,\rho_t^{\alpha,\sigma})(X_t^\alpha), \partial_x\partial_\mu v(t,\rho_t^{\alpha,\sigma})(X_t^\alpha),\alpha_t,\Sigma_t) \big],  
\enq
with $H$ the function defined on $\R\times\R\times\R^d\times\S_{>+}^{d}$ by
\beq \label{defH}
H(p,M,a,\Sigma) &=& p a\trans b +  \frac{1}{2} M a\trans\Sigma a. 
\enq

We state some easy properties for the function $H$, which allows us to introduce some useful notations.

\begin{Lemma} \label{propH}
For all $(p,M)$ $\in$ $\R\times (0,\infty)$, $a$ $\in$ $A$, we have
\beqs
\sup_{\Sigma\in\Gamma} H(p,M,a,\Sigma) &=& H(p,M,a,\hat\Sigma(a))  \; < \; \infty, \; 
\;\;\; \mbox{ with } \; \hat\Sigma(a) \in {\rm arg}\max_{\Sigma\in \Gamma} a\trans\Sigma a. 
\enqs
There exists a measurable function $(p,M)$ $\in$ $\R\times (0,\infty)$ $\mapsto$ $a^*(p,M)$ $\in$ $A$ such that 
\beq \label{relH}
H^*(p,M) \; := \;  \inf_{a\in A}  \sup_{\Sigma\in\Gamma} H(p,M,a,\Sigma)  &=& \sup_{\Sigma\in\Gamma}  H(p,M,a^*(p,M),\Sigma). 
\enq
\end{Lemma}
{\bf Proof.} For fixed $(p,M)$ $\in$ $\R\times (0,\infty)$, $a$ $\in$ $A$, it is clear that the continuous function  $\Sigma$ $\mapsto$ $H(p,M,a,\Sigma)$ 
attains its maximum on the compact set $\Gamma$ at some point $\hat\Sigma(a)$ given by   $\hat\Sigma(a) \in {\rm arg}\max_{\Sigma\in \Gamma} a\trans\Sigma a$, 
from the expression of $H$, hence not depending on  $(p,M)$.   By convexity of the function $a$ $\mapsto$ $|a|^2$, it is clear that the function $a$ $\in$ $A$ $\mapsto$ $\bar H(p,M,a)$ $:=$ $\sup_{\Sigma\in\Gamma} H(p,M,a,\Sigma)$ is also convex. Moreover, since 
$\bar H(p,M,a)$  $\geq$ $p a\trans b +  \frac{1}{2} M a\trans\underline{\Sigma}a$, with $\underline{\Sigma}$ positive definite, 
we see that $\bar H(p,M,a)$ goes to infinity when $|a|$ goes to infinity. It follows that $a$ $\mapsto$ $\bar H(p,M,a)$  attains its infimum on the closed convex set $A$ at some $a^*(p,M)$ which can be chosen measurable by continuity of $H$ and Carath\'eodory-type measurable selection theorem, see e.g. \cite{wag80}.  
\ep

\vspace{2mm}

We can now state an analytic verification theorem for the robust mean-variance portfolio selection problem, which provides a characterization of the optimal portfolio strategy. 

\begin{Theorem} \label{theoverif} (Verification theorem)

\noindent  Let  $v$ be  a smooth function on  $[0,T]\times\Pc_{_2}(\R)$ satisfying $\partial_x\partial_\mu v(t,\mu)(x)$ $>$ $0$ for all $(t,x,\mu)$ $\in$ 
$[0,T)\times\R\times\Pc_{_2}(\R)$, and suppose that $v$ is solution to the Bellman-Isaacs partial differential equation (PDE): 
\begin{equation} \label{PDE}
\left\{
\begin{array}{rcl}
\partial_t v(t,\mu) + \Int_\R H^*\big(\partial_\mu v(t,\mu)(x),\partial_x\partial_\mu v(t,\mu)(x) \big) \mu(dx) &=& 0, \;\;\; (t,\mu) \in [0,T)\times\Pc_{_2}(\R) \\
v(T,\mu) &=&  \lambda {\rm Var}(\mu) - \bar\mu, \;\;\;\; \mu \in  \Pc_{_2}(\R),
\end{array}
\right.
\end{equation}
s.t. the function $(x,\mu)$ $\in$ $\R\times\Pc_{_2}(\R)$ $\mapsto$ $\hat a(t,x,\mu)$ $:=$ $a^*(\partial_\mu v(t,\mu)(x),\partial_x\partial_\mu v(t,\mu)(x))$ is Lip\-schitz, for any $t$ $\in$ $[0,T]$, and $\int_0^T |\hat a(t,0,\delta_0)|^2 dt$ $<$ $\infty$.  
For any $\Sigma$ $\in$ $\Vc_\Theta$, denote by $X^{\P^\sigma}$ the solution to the McKean-Vlasov SDE under $\P^\sigma$:
\beq \label{SDEMK}
dX_t &=& \hat a(t,X_t,\P^\sigma_{_{X_t}}) [ b dt + \sigma_t dW_t^{\sigma}], \;\;\; 0 \leq t \leq T, \; X_0 = x_0, \; \P^\sigma-\mbox{p.s.} 
\enq
and suppose that the family of processes $\{X^{\P^\sigma}, \Sigma\in\Vc_\Theta\}$ 
can be aggregated into a $\Pc^\Theta$-quasi surely aggregated solution, i.e. there exists $X^*$ s.t.
\beqs
X_t^* &=& X_t^{\P^\sigma},  \;\;\; 0 \leq t \leq T, \; \P^\sigma-\mbox{p.s.}, \; \forall \Sigma\in\Vc_\Theta. 
\enqs
Then, the family of processes $\{\hat a(t,X_t^{\P^\sigma},\P^\sigma_{_{X_t^{\P^\sigma}}}), 0\leq t \leq T, \Sigma\in\Vc_\Theta\}$  can also  be aggregated, i.e. there exists  a process $\alpha^*$ s.t. 
\beq \label{defalpha*}
\alpha_t^* &=& \hat a(t,X_t^{\P^\sigma},\P^\sigma_{_{X_t^{\P^\sigma}}})
\;\;\; 0 \leq t \leq T, \; \P^\sigma-\mbox{p.s.}, \; \forall \Sigma\in\Vc_\Theta,  
\enq
and the process $\alpha^*$ defines a portfolio strategy in $\Ac$, which 
is optimal  for \reff{robustMV}, i.e. $V_0$ $=$ $J_{wc}(\alpha^*)$, and we have $V_0$ $=$ $v(0,\delta_{x_0})$.  
\end{Theorem}

\begin{Remark} \label{remtheoverif}
{\rm {\bf 1.} In standard  stochastic control problem where the criterion involves linear functional of the law of the state process, we look for a value function $v(t,\mu)$, which is also linear in $\mu$, 
hence of the form $v(t,\mu)$ $=$ $\int w(t,x) \mu(dx)$ for some smooth function $w$ on $[0,T]\times\R$ solution to the standard Hamilton-Jacobi-Bellman-Isaacs equation. 
In this case, $\partial_x\partial_\mu v(t,\mu)(x)$ $=$ $D_x^2 w(t,x)$, and the above condition in the verification theorem: $\partial_x\partial_\mu v(t,\mu)(x)$ $>$ $0$ for all $(t,x,\mu)$, simply 
means that we look for a convex function $w$, which usually follows from the convexity of the terminal cost and the linear dynamics of the wealth process.  Here for the mean-variance criterion, 
the condition $\partial_x\partial_\mu v(t,\mu)(x)$ $>$ $0$ is related to the positivity of the variance penalization parameter $\lambda$, see \reff{derivv}.   

\noindent {\bf 2.} For fixed $\Sigma$ $\in$ $\Vc_\Theta$,  
the existence and uniqueness of a $\P^\sigma$-solution $X^{\P^\sigma}$  to the McKean-Vlasov SDE \reff{SDEMK} 
under the Lipschitz condition on $\hat a$ and the square-integrability condition of $\hat a(.,0,\delta_0)$ follows from standard arguments (recall that $\Sigma$ $\in$ $\Vc_\Theta$ is bounded) 
as in \cite{sni89} or \cite{jouetal08}, and we have the estimate: 
\beq \label{integXuni}
\E_\sigma \big[ \sup_{0\leq t\leq T} |X_t^{\P^\sigma}|^2 \big] & \leq & C( 1 + \int_0^T |\hat a(t,0,\rho_t^*)|^2 dt \big) \; <  \; \infty, 
\enq
for some positive constant $C$ depending on the Lipschitz condition on the function $x$ $\mapsto$ $\hat a(t,x,\rho_t^*)$, and independent of $\Sigma$. 
The key assumption in the above verification theorem is the fact one can aggregate the family of processes $\{X^{\P^\sigma}, \Sigma\in\Vc_\Theta\}$ in order to define a universal process 
$X^*$ defined $\Pc^\Theta$-quasi surely. This point is discussed more precisely in the next section,  where it is shown that the aggregation condition is satisfied when prior probability measures 
are related to uncertainty on covariance matrix  (see Theorem \ref{theomain2}), but not in general on drift uncertainty (see Remark \ref{remdrift}). Once this aggregation condition is satisfied, we notice that the $i$-th component of the $\R^d$-valued process $\{\hat a(t,X_t^{\P^\sigma},\P^\sigma_{_{X_t^{\P^\sigma}}}), 0\leq t \leq T\}$ is obtained  as the Radon-Nikodym derivative
\beqs
\hat a^i(t,X_t^{\P^\sigma},\P^\sigma_{_{X_t^{\P^\sigma}}}) &=& \frac{ d<X^*,B^i>_t}{d<B^i>_t}, \;\;\; 0 \leq t \leq T, \; \P^\sigma-\mbox{p.s.}, \; \forall \Sigma\in\Vc_\Theta, 
\enqs
where $<X^*,B^i>$ is the quadratic covariation (covariance) process associated to $X^*$ and $B^i$, which is defined $\Pc^\Theta$-quasi surely.  Therefore, the family of processes $\{\hat a(t,X_t^{\P^\sigma},\P^\sigma_{_{X_t^{\P^\sigma}}}), 0\leq t \leq T, \Sigma\in\Vc_\Theta\}$  can   be aggregated into $\alpha^*$ as in \reff{defalpha*}, and 
we easily see from \reff{integXuni} that $\alpha^*$ satisfies the integrability condition \reff{integalpha}, hence lies in $\Ac$. By construction, we then see that $X^*$ $=$ 
$X^{\alpha^*}$ the associated (self-financing) wealth process, and the remaining point in the verification theorem is  to check that $\alpha^*$ is optimal, as proved below.  
}
\ep
\end{Remark}

\noindent {\bf Proof of Theorem \ref{theoverif}.}   
It suffices to check that the family of (deterministic) processes $V_t^{\alpha,\sigma}$ $=$ $v(t,\rho_t^{\alpha,\sigma})$, $0\leq t\leq T$, with $v$ solution to the PDE \reff{PDE}, satisfies 
the conditions of the optimality principle with $\alpha^*$. 
Condition (i) is already satisfied and in view of \reff{dV},  it suffices to check that (ii) for all $\alpha$ $\in$ $\Ac$, there exists 
$\bar\Sigma$ depending on $\alpha$ $\in$ $\Vc_\theta$ s.t. $D_t^{\alpha,\bar\sigma}$ $\geq$ $0$, $0\leq t\leq T$,  and 
(iii) $D_t^{\alpha^*,\sigma}$ $\leq$ $0$, $0\leq t\leq T$, for all  $\Sigma$ $\in$ $\Vc_\Theta$, hold true. 
Given $\alpha$ $\in$ $\Ac$, consider the process $\bar\Sigma$ $\in$ $\Vc_\Theta$ defined by $\bar\Sigma_t$ $=$ $\hat\Sigma(\alpha_t)$, $0\leq t\leq T$, 
where $\hat\Sigma(.)$ is defined in Lemma \ref{propH}.  Recalling the expression of 
$D^{\alpha,\bar\sigma}$ in \reff{defD}, we have for all  $t$ $\in$ $[0,T]$, 
\beqs
D_t^{\alpha,\bar\sigma} &=&  
\E_{\bar\sigma} \big[  \partial_t v(t,\rho_t^{\alpha,\bar\sigma})  + 
H(\partial_\mu v(t,\rho_t^{\alpha,\bar\sigma})(X_t^\alpha), \partial_x\partial_\mu v(t,\rho_t^{\alpha})(X_t^\alpha),\alpha_t,\hat\Sigma(\alpha_t)) \big] \\
&=&  \E_{\bar\sigma} \big[  \partial_t v(t,\rho_t^{\alpha,\bar\sigma})  + \sup_{\gamma\in\Gamma}
H(\partial_\mu v(t,\rho_t^{\alpha,\bar\sigma})(X_t^\alpha), \partial_x\partial_\mu v(t,\rho_t^{\alpha,\bar\sigma})(X_t^\alpha),\alpha_t,\gamma) \big] \\
& \geq &  \E_{\bar\sigma} \big[  \partial_t v(t,\rho_t^{\alpha,\bar\sigma})  +  
H^*(\partial_\mu v(t,\rho_t^{\alpha,\bar\sigma})(X_t^\alpha), \partial_x\partial_\mu v(t,\rho_t^{\alpha,\bar\sigma})(X_t^\alpha)) \big] \; = \; 0, 
\enqs
where the second equality comes from the definition of $\bar\Sigma_t$ $=$ $\hat\Sigma(\alpha_t)$, the inequality $\geq$ from the fact that 
$H^*(p,M)$ $\leq$  $\sup_{\gamma\in\Gamma} H(p,M,a,\gamma)$ for all $a$ $\in$ $A$, and the last equality $=$ $0$ from the PDE \reff{PDE} satisfied by 
$v$ at point $(t,\rho_t^{\alpha,\bar\sigma})$ and recalling that  $\rho^{\alpha,\bar\sigma}$ is the law of $X_t^\alpha$ under $\P^{\bar\sigma}$. This proves the condition (ii). On the other  hand, 
let us consider the universal process $\alpha^*$ $\in$ $\Ac$ defined in \reff{defalpha*}. We then have for all $\Sigma$ $\in$ $\Vc_\Theta$, and $t$ $\in$ $[0,T]$, 
\beqs
D_t^{\alpha^*,\sigma} &=&  
 \E_{\sigma} \big[ \partial_t v(t,\rho_t^{\alpha^*,\sigma})  
+  H(\partial_\mu v(t,\rho_t^{\alpha^*,\sigma})(X_t^*), \partial_x\partial_\mu v(t,\rho_t^{\alpha^*,\sigma})(X_t^*),\alpha_t^*,\Sigma_t) \big] \\
& \leq &   \E_{\sigma} \big[ \partial_t v(t,\rho_t^{\alpha^*,\sigma})  
+  \sup_{\gamma\in\Gamma} H(\partial_\mu v(t,\rho_t^{\alpha^*,\sigma})(X_t^*), \partial_x\partial_\mu v(t,\rho_t^{\alpha^*,\sigma})(X_t^*),\alpha_t^*,\gamma) \big] \\
&=& \E_{\sigma} \big[ \partial_t v(t,\rho_t^{\alpha^*,\sigma})  + 
H^*(\partial_\mu v(t,\rho_t^{\alpha^*,\sigma})(X_t^*), \partial_x\partial_\mu v(t,\rho_t^{\alpha^*,\sigma})(X_t^*)) \big] \; = \; 0, 
\enqs
where the second equality follows from the definition of $\alpha^*$ and relation \reff{relH}. This proves condition (iii), and ends the proof of this theorem.
\ep

\section{Explicit solutions}  \label{secexpli}

\setcounter{equation}{0}
\setcounter{Assumption}{0} \setcounter{Theorem}{0}
\setcounter{Proposition}{0} \setcounter{Corollary}{0}
\setcounter{Lemma}{0} \setcounter{Definition}{0}
\setcounter{Remark}{0}

We provide in this section explicit solutions to the Bellman-Isaacs PDE \reff{PDE},  hence to the robust mean-variance portfolio selection problem \reff{robustMV},   when $A$ $=$ $\R^d$, and 
for a  class of prior models $\Gamma$ on the covariance matrix satisfying a concavity assumption. Recall our parametrization of the covariance matrix: there is some convex set 
$\Theta$  $\subset$ $\R^q$, and a measurable function $\gamma$ $:$ $\R^q$ $\rightarrow$ $\S_{>+}^{d}$ s.t. any $\Sigma$ $\in$ $\Gamma$ $=$ $\Gamma(\Theta)$ is in the form $\Sigma$ $=$ $\gamma(\theta)$ for some $\theta$ in $\Theta$. We shall assume that

\vspace{2mm}

{\bf (IC)} $A$ $=$ $\R^d$ and $\gamma$ $:$ $\R^q$ $\rightarrow$ $\S_{>+}^{d}$ is concave\footnote{We use the partial ordering $\preceq$ on the set of $d\times d$-symmetric matrices: 
$M$ $\preceq$ $N$ $\Leftrightarrow$ $N-M$ is  positive semi-definite $\Leftrightarrow$ $a\trans(N-M)a$ $\geq$ $0$ for all $a$ $\in$ $\R^d$.} 
 on $\Theta$, i.e. for all $\theta_1$, $\theta_2$ $\in$ $\Theta$, 
\beq \label{IC}
\frac{1}{2} \big( \gamma(\theta_1) + \gamma(\theta_2) \big) &  \preceq &  \gamma\big(\frac{1}{2}(\theta_1 +\theta_2)\big).  
\enq
Notice that this assumption is trivially satisfied  in the Examples 1 and 2 of uncertain volatilities and ambiguous correlation detailed in Section \ref{secprob} where we have actually equality in \reff{IC}.

\vspace{2mm}

Let us denote by $R$ the (square) risk premium function: 
\beq \label{riskpremium}
R(\theta) &:=& b\trans\gamma(\theta)^{-1}b, \;\;\; \theta \in \Theta. 
\enq

The next Lemma provides a key result on the Hamiltonian function $H$ in \reff{defH}, which will be useful  for the  elucidation of our problem.


\begin{Lemma} \label{lemIC} 
Let condition {\bf (IC)} hold. Then, for all $p$ $\in$ $\R$, $M$ $>$ $0$, we have 
\beq
H^*(p,M) &= & - \frac{1}{2} \frac{p^2}{M} b\trans (\Sigma^*)^{-1}b  \label{expliH*} \\
&=& H(p,M,a^*(p,M),\Sigma^*) \nonumber
\enq
where $\Sigma^*$ $=$ $\gamma(\theta^*)$ is a constant in $\Gamma$ $=$ $\Gamma(\Theta)$ defined by 
\beq \label{sig*}
\Sigma^* & \in  & 
{\rm arg}\min_{\Sigma\in\Gamma} \big[  b\trans \Sigma^{-1} b \big], \; \mbox{ i.e. } \;  \theta^* \; \in \;   {\rm arg}\min_{\theta\in\Theta} R(\theta). 
\enq
Moreover, the pair $(a^*,\Sigma^*)$ is a saddle-point for $H$ i.e.  for all  $p$ $\in$ $\R$, $M$ $>$ $0$,
\begin{equation} \label{saddle}
\left\{
\begin{array}{rcl}
H(p,M,a^*(p,M),\Sigma) & \leq &  H(p,M,a^*(p,M),\Sigma^*) \; = \; H^*(p,M), \;\;\; \forall \Sigma \in \Gamma, \\
H(p,M,a,\Sigma^*) & \geq &  H(p,M,a^*(p,M),\Sigma^*)\; = \; H^*(p,M), \;\;\; \forall a \in \R^d,  
\end{array}
\right.
\end{equation}
and $a^*$ is explicitly given by: 
\beq \label{explia*}
a^*(p,M) &=& -  \frac{p}{M}(\Sigma^*)^{-1}b.  
\enq
\end{Lemma} 
{\bf Proof.} Denote by $\tilde H$ the function defined on $\R\times\R\times\R^d\times\Theta$ by 
\beqs
\tilde H(p,M,a,\theta) & := & H(p,M,a,\gamma(\theta)) \;  = \; p a\trans b +  \frac{1}{2} M a\trans\gamma(\theta) a.
\enqs
Under the concavity assumption of $\gamma$ in {\bf (IC)}, we clearly see that  for fixed $(p,M)$ $\in$ $\R\times(0,\infty)$, the function $\tilde H(p,M,.,.)$ is convex in $a$ $\in$ $\R^d$, and  concave in $\theta$ lying in the convex-compact set $\Theta$. By the min-max theorem (see e.g. Theorem 45.8 in \cite{stra85}), we then get the so-called Isaacs relation: 
\beqs
\inf_{a \in \R^d} \sup_{\theta \in \Theta} \tilde H(p,M,a,\theta) & = &   \sup_{\theta \in \Theta}  \inf_{a \in \R^d}  \tilde H(p,M,a,\theta), \\
 \mbox{ i.e. } \;\;\;  \inf_{a \in \R^d} \sup_{\Sigma \in \Gamma} H(p,M,a,\Sigma) & = &   \sup_{\Sigma \in \Gamma}  \inf_{a \in \R^d}  H(p,M,a,\Sigma).  
\enqs 
By square completion, we  can rewrite the function $H$  as: 
\beq \label{squareH}
H(p,M,a,\Sigma) &=& \frac{M}{2} \big( a + \frac{p}{M} \Sigma^{-1}b \big)\trans\Sigma(a + \frac{p}{M} \Sigma^{-1}b \big) - \frac{1}{2} \frac{p^2}{M} b\trans \Sigma^{-1}b,
\enq
from which we get
\beq \label{infH}
\inf_{a \in \R^d}  H(p,M,a,\Sigma) &=&  H(p,M,\bar a(p,M,\Sigma),\Sigma) \; = \;    - \frac{1}{2} \frac{p^2}{M} b\trans \Sigma^{-1}b,
\enq
where we set: $\bar a(p,M,\Sigma)$ $:=$ $-\frac{p}{M}\Sigma^{-1}b$, and then the explicit expression of $H^*(p,M)$ 
\beqs
H^*(p,M) \; = \; \sup_{\Sigma \in \Gamma} \inf_{a \in \R^d} H(p,M,a,\Sigma) &=& -  \frac{1}{2} \frac{p^2}{M} \inf_{\Sigma \in \Gamma} b\trans \Sigma^{-1}b
\; = \;  - \frac{1}{2} \frac{p^2}{M} b\trans (\Sigma^*)^{-1}b.
\enqs
Let us now check the saddle-point property of $(a^*,\Sigma^*)$. By definition of $a^*(p,M)$, we have 
\beqs
\sup_{\Sigma\in\Gamma} H(p,M,a^*(p,M),\Sigma) &=&  \inf_{a \in \R^d} \sup_{\Sigma \in \Gamma} H(p,M,a,\Sigma)  \\
&=&   \sup_{\Sigma \in \Gamma}  \inf_{a \in \R^d}  H(p,M,a,\Sigma) \;= \; H^*(p,M) \\
&=&  \inf_{a \in \R^d}  H(p,M,a,\Sigma^*) \; \leq \;  H(p,M,a,\Sigma^*), \;\;\; \forall a \in \R^d,  
\enqs
where we used in the second equality Isaacs condition, and noticed  in the last equality  that $\Sigma^*$ attains the supremum of 
$\Sigma$ $\mapsto$  $\inf_{a \in \R^d}  H(p,M,a,\Sigma)$ by \reff{infH}. We then deduce
\beqs
H(p,M,a^*(p,M),\Sigma^*) \; \leq \;  \sup_{\Sigma\in\Gamma} H(p,M,a^*(p,M),\Sigma) & = & H^*(p,M)  \\
& \leq &  H(p,M,a,\Sigma^*), \;\;\; \forall a \in \R^d,
\enqs
which shows the second inequality  in  \reff{saddle}. Similarly, we have 
\beqs
\inf_{a \in A } H(p,M,a,\Sigma^*) &=&   \sup_{\Sigma \in \Gamma} \inf_{a \in \R^d}  H(p,M,a,\Sigma)  \\
&=&    \inf_{a \in \R^d} \sup_{\Sigma \in \Gamma}    H(p,M,a,\Sigma) \;= \; H^*(p,M) \\
&=&    \sup_{\Sigma \in \Gamma}  H(p,M,a^*(p,M),\Sigma) \; \geq \;  H(p,M,a^*(p,M),\Sigma), \;\;\; \forall \Sigma \in \Gamma,
\enqs
which implies that
\beqs
H(p,M,a^*(p,M),\Sigma^*) \; \geq \;  \inf_{a \in\R^d} H(p,M,a,\Sigma^*) & = & H^*(p,M)  \\
& \geq &  H(p,M,a^*(p,M),\Sigma), \;\;\; \forall \Sigma \in \Gamma.
\enqs
This proves the first inequality in  \reff{saddle}, hence the saddle-point property, and also that $H^*(p,M)$ $=$ 
$H(p,M,a^*(p,M),\Sigma^*)$. 

On the other hand, by applying  relation \reff{infH} to  $\Sigma$ $=$ $\Sigma^*$, we have
\beqs
H^*(p,M) \; = \; H(p,M,\bar a(p,M,\Sigma^*),\Sigma^*),
\enqs
which combined with the saddle-point property of $(a^*,\Sigma)$ shows that: $H(p,M,a^*(p,M),\Sigma^*)$ $=$ $H(p,M,\bar a(p,M,\Sigma^*),\Sigma^*)$ $=$ $H^*(p,M)$, and then from the expression \reff{squareH} of $H$: 
\beqs
 \frac{M}{2} \big( a^*(p,M) -  \bar a(p,M,\Sigma^*) \big)\trans\Sigma^*(a -  \bar a(p,M,\Sigma^*)  \big)  +  H^*(p,M) &=& H^*(p,M). 
\enqs
This proves that $a^*(p,M)$ $=$ $\bar a(p,M,\Sigma^*)$, i.e. the expression \reff{explia*}. 
\ep

\vspace{1mm}

\begin{Remark}
{\rm Under condition {\bf (IC)}, and if the conditions of the verification theorem \ref{theoverif} are satisfied with a solution $v$ to the Bellman-Isaacs PDE \reff{PDE}, and an optimal feedback control $\alpha^*$, then we see from the saddle-point relation \reff{saddle}, that the drift $D_t^{\alpha,\sigma}$ of the deterministic process $V_t^{\alpha,\sigma}$ $=$ $v(t,\rho_t^{\alpha,\sigma})$ satisfies for all $\alpha$ $\in$ $\Ac$, $\Sigma$ $\in$ $\Vc_\Theta$, 
\beqs
D_t^{\alpha,\sigma^*} & \geq \; D_t^{\alpha^*,\sigma^*} \; = \; 0 & \geq \; D_t^{\alpha^*,\sigma}, \;\;\; 0 \leq t \leq T, \; a.s.,
\enqs
where $\sigma^*$ $=$ $(\Sigma^*)^{1\over 2}$. This means that the process (i) $V_t^{\alpha,\sigma^*}$ is nondecreasing for all $\alpha$ $\in$ $\Ac$,  (ii) the process $V_t^{\alpha^*,\sigma}$ is nonincreasing for all $\Sigma$ $\in$ $\Vc_\Theta$, from which we easily deduce the min-max property:
\beqs
V_0 \;  = \; v(0,\delta_{_{x_0}})  &=& \inf_{\alpha\in\Ac} \sup_{\Sigma\in\Vc_\Theta} J(\alpha,\sigma) 
\; = \;  \sup_{\Sigma\in\Vc_\Theta} \inf_{\alpha\in\Ac} J(\alpha,\sigma) \; = \; J(\alpha^*,\sigma^*). 
\enqs
This shows in particular that $\sigma^*$, which is a constant explicitly computed from \reff{sig*}, i.e. minimizing the risk premium,  
is an optimal worst-case volatility for the robust mean-variance problem. 
}
\ep
\end{Remark}

\vspace{2mm}

\begin{Proposition} \label{proexpli}
Assume that  {\bf (IC)} holds. Then, the function defined on $[0,T]\times\Pc_{_2}(\R)$ by
\beq \label{expliv}
v(t,\mu) &=& K(t){\rm Var}(\mu)  - \bar \mu + \chi(t), 
\enq
with 
\begin{equation} \label{KY}
\left\{
\begin{array}{ccl}
K(t) &=& \lambda \exp\big( -   R^* (T-t) \big) \\
\chi(t) &=& - \frac{1}{4\lambda} \Big[ \exp\big(   R^*  (T-t) \big) - 1 \Big], \;\;\; 0 \leq t\leq T, \\
R^* &=& b\trans(\Sigma^*)^{-1}b,  
\end{array}
\right.
\end{equation}
is solution to the Bellman-Isaacs PDE \reff{PDE}. 
\end{Proposition}
{\bf Proof.}
We look for a function solution to \reff{PDE} in the form: 
\beq \label{vquadra}
v(t,\mu) &=& K(t){\rm Var}(\mu)  + Y(t) \bar \mu + \chi(t),
\enq
for some continuously differentiable functions $K$ $>$ $0$, $Y$ and $\chi$ on $[0,T]$.  Such function is smooth and we have
\beqs
\partial_\mu v(t,\mu)(x) \; = \;  2 K(t) (x - \bar \mu) + Y(t),  & & \partial_x \partial_\mu v(t,\mu)(x) \; = \; 2 K(t) > 0,  
\enqs
From the expression of $H^*$ in \reff{expliH*}, we then get
\beqs
 & &  \partial_t v(t,\mu) + \Int_\R H^*\big(\partial_\mu v(t,\mu)(x),\partial_x\partial_\mu v(t,\mu)(x) \big) \mu(dx) \\
 &=& \dot K(t){\rm Var}(\mu) +  \dot Y(t) \bar \mu +  \dot \chi(t) \\
 & & \;\;\; - \;  \frac{1}{2} b\trans(\Sigma^*)^{-1}b \int \frac{4 K(t)^2 (x-\bar\mu)^2 + Y(t)^2 + 4 K(t)Y(t)(x-\bar\mu)}{2K(t)} \mu(dx) \\
 &=& \big[ \dot K(t) - b\trans(\Sigma^*)^{-1}b K(t) \big] {\rm Var}(\mu) + \dot Y(t) \bar \mu +  \dot \chi(t) 
 - \frac{1}{4} b\trans(\Sigma^*)^{-1}b \frac{Y(t)^2}{K(t)}. 
\enqs
It follows that $v$ in \reff{vquadra} satisfies the Bellman-Isaacs PDE \reff{PDE} iff $K$, $Y$ and $\chi$ satisfy the system of ordinary differential equations: 
\beqs
\dot K(t) - b\trans(\Sigma^*)^{-1}b K(t) &=& 0, \;\;\; K(T) \; = \; \lambda  \\
\dot Y(t) &=& 0, \;\;\; Y(T) \; = \; -1 \\
 \dot \chi(t) - \frac{1}{4} b\trans(\Sigma^*)^{-1}b \frac{Y(t)^2}{K(t)} &=& 0, \;\;\; \chi(T) \; = \; 0, 
\enqs
which leads to  the explicit solution $Y$ $=$ $-1$,  $K$, $\chi$ as in  \reff{KY}. 
\ep

\vspace{2mm}

We can now provide a complete and explicit  resolution of the robust mean-variance problem for a general class of  covariance matrix uncertainty model satisfying {\bf (IC)}. 

\begin{Theorem} \label{theomain2}
Let condition {\bf (IC)} hold. There exists an optimal robust mean-variance strategy solution to \reff{robustMV}, and given explicitly by 
\beq\label{explialpha*gen}
\alpha_t^* &=&  \Big[ x_0 + \frac{1}{2\lambda} \exp\big(   R^*  T  \big) - X_t^* \Big](\Sigma^*)^{-1} b, \;\;\; 0 \leq t \leq T, \;\; \Pc^\Theta-q.s. 
\enq
where $R^*$ $=$ $b\trans(\Sigma^*)^{-1} b$ is the minimal risk premium corresponding to the worst-case covariance matrix parameter $\Sigma^*$, 
and with an optimal corresponding wealth process $X^*$, whose terminal return under any $\P^\sigma$, $\Sigma$ $\in$ $\Vc_\Theta$ is given by:
\beq \label{returnopt}
\E_\sigma[X_T^*] &=& x_0 + \frac{1}{2\lambda}   \Big[\exp\big(R^*  T \big) -1 \Big]. 
\enq
Moreover,  the optimal cost is given by 
\beq \label{Vopti}
V_0 \;= \;  v(0,\delta_{x_0}) &=& - \frac{1}{4\lambda}  \big[ \exp\big(   R^* T  \big)  -1 \big]  - x_0. 
\enq
\end{Theorem}
{\bf Proof.} Let us consider the  function $v(t,\mu)$ in \reff{expliv}, which satisfies the Bellman-Isaacs PDE \reff{PDE}. For this smooth function on $[0,T]\times\Pc_{_2}(\R)$, we have 
\beq \label{derivv}
\partial_\mu v(t,\mu)(x) \; = \;  2 K(t) (x - \bar \mu) - 1,  & & \partial_x \partial_\mu v(t,\mu)(x) \; = \; 2 K(t) > 0,   
\enq
with $K$ as in \reff{KY}. From the expression of $a^*$ in \reff{explia*}, the candidate $\hat a(t,x,\mu)$ for the optimal feedback control in the verification Theorem \ref{theoverif} is then equal to:
\beqs
\hat a(t,x,\mu) & := &  a^*(\partial_\mu v(t,\mu)(x),\partial_x\partial_\mu v(t,\mu)(x)) \\
&=& -  \Big[ x- \bar\mu  -   \frac{1}{2 K(t)}  \Big](\Sigma^*)^{-1} b,
\enqs
which is clearly Lipschitz in $(x,\mu)$. 
The solution $X^{\P^\sigma}$ to the McKean-Vlasov SDE \reff{SDEMK} under $\P^\sigma$ $\in$ $\Pc^\Theta$, is thus governed by
\beqs
dX_t^{\P^\sigma} &=&  - \Big[  X_t^{\P^\sigma} - \E_\sigma[X_t^{\P^\sigma}] -    \frac{1}{2 K(t)} \Big] b\trans (\Sigma^*)^{-1}
[ b dt + \sigma_t dW_t^{\sigma}], \;\; 0 \leq t \leq T,  \; \P^\sigma-\mbox{p.s.},  
\enqs
which yields by taking expectation under $\P^\sigma$:
\beqs
d \E_\sigma[X_t^{\P^\sigma}]  &=&  \frac{R^*}{2 K(t)} dt, 
\enqs
and thus 
\beqs
\E_\sigma[X_t^{\P^\sigma}] &=& x_0 +   \int_0^t  \frac{R^*}{2 K(s)} ds,  \;\;\;  0 \leq t\leq T.
\enqs
The crucial observation is that this expectation  does not depend on $\Sigma$ $\in$ $\Vc_\Theta$. 
Plugging into the SDE \reff{SDEMK}, this can be rewritten as (we now simply write $X_t$ $=$ $X_t^{\P^\sigma}$ to alleviate notations): 
\beqs
dX_t &=&  - \Big[  X_t - x_0 -   \int_0^t  \frac{R^*}{2 K(s)} ds -    \frac{1}{2 K(t)} \Big] b\trans (\Sigma^*)^{-1}
[ b dt + dB_t ], \;\; 0 \leq t \leq T,  \; \Pc^\Theta-\mbox{q.s.}.
\enqs 
This is now a standard SDE under $\Pc^\Theta$, and we know from Proposition 6.10 in \cite{sonetal11} that there exists a $\Pc^\Theta$ quasi surely aggregated solution $X^*$, i.e. 
$X^*$ $=$ $X^{\P^\sigma}$, $\P^\sigma$-p.s, for all  $\Sigma$ $\in$ $\Vc_\Theta$. Consequently, the family of processes $\{\hat a(t,X_t^{\P^\sigma},\P^\sigma_{_{X_t^{\P^\sigma}}}), 0\leq t \leq T, \Sigma\in\Vc_\Theta\}$ can be aggregated into a $\Pc^\Theta$-q.s. defined by 
\beqs 
\alpha_t^* &=& \hat a(t,X_t^{\P^\sigma},\P^\sigma_{_{X_t^{\P^\sigma}}}), \;\;\; 0 \leq t \leq T, \; \P^\sigma-\mbox{p.s.}, \; \forall \Sigma\in\Vc_\Theta \\
&=&    -  \Big[ X_t^* - x_0 -   \int_0^t  \frac{R^*}{2 K(s)} ds -    \frac{1}{2 K(t)} \Big](\Sigma^*)^{-1} b, \;\;\; 0 \leq t \leq T, \;\; \Pc^\Theta-\mbox{q.s.},
\enqs
which gives from \reff{KY}  the expression in \reff{explialpha*gen}.  We conclude from the verification Theorem \ref{theoverif} that $\alpha^*$ is an optimal solution to \reff{robustMV}, and the optimal cost is equal to $V_0$ $=$ $v(0,\delta_{x_0})$, hence given by \reff{Vopti} from the explicit form of $v$ in \reff{expliv}. 
\ep

\begin{Remark} \label{remgen}
{\rm Although the original robust mean-variance problem is {\it a priori} a complex and non standard stochastic differential game problem, the message of the main result in Theorem \ref{theomain2} is quite simple with an intuitive interpretation. It says that the resolution of this problem can be reduced into two steps: 
first, we determine the worst-case scenario, and the remarkable point is that it corresponds to a constant covariance matrix $\Sigma^*$ 
obtained by the minimization \reff{sig*} of the risk premium. This constant is directly computed  from the inputs of the model: the instantaneous return $b$ (assumed to be known), and the function 
$\gamma$ parametrizing the uncertainty on the covariance matrix  of the assets (we shall give in the sequel some examples for explicit computations of $\Sigma^*$).  
Secondly, we obtain the optimal mean-variance strategy as in the Black-Scholes model with instantaneous return $b$ and covariance matrix $\Sigma^*$, whose expression has been derived in  \cite{zholi00}, and that we recover here by a different approach as a particular case when there is no uncertainty on the model. 
}
\ep
\end{Remark}

\begin{Remark} \label{remdrift} (About drift uncertainty)
{\rm 

\noindent Let us discuss the case when there is ambiguity on the drift of the $d$ risky assets (but with known covariance matrix $\Sigma$ for simplicity). This is modeled by considering that the drift process $b$ $=$ $(b_t)_t$ $\in$ $\Vc_\Theta$ is an unobservable process, which is only known to be valued in a given convex set $\Theta$ of $\R^d$. The Hamiltonian function for the corresponding robust optimization problem is then given by (by abuse of notation, we keep the same notation $H$ as in the case of uncertain covariance matrix): 
\beqs
H(p,M,a,\theta) &=& pa\trans\theta + \frac{1}{2}M a\trans\Sigma a, \;\;\; (p,M,a,\theta) \in \R\times (0,\infty)\times\R^d\times\Theta.  
\enqs
By similar arguments as in Lemma \ref{lemIC}, for fixed $(p,M)$ $\in$ $\R\times (0,\infty)$, there is a saddle-point $(a^*(p,M),\theta^*)$ for $H(p,M,.,.)$ given by 
\beqs
a^*(p,M) \; = \;  -  \frac{p}{M}\Sigma^{-1}\theta^*, & &   \theta^* \; \in \;  {\rm arg}\min_{\theta\in\Theta} \big[  \theta\trans \Sigma^{-1} \theta \big]. 
\enqs
Then, similarly as in Proposition \ref{proexpli}, we find that $v$ given in \reff{expliv}-\reff{KY} is solution to the associated Bellman-Isaacs PDE for the robust mean-variance problem, where 
the "worst-case" risk premium $R^*$ is now given by
\beqs
R^* &=& (\theta^*)\trans \Sigma^{-1} \theta^*. 
\enqs
Following arguments as in the verification Theorem \ref{theoverif}, this leads to a candidate for the optimal feedback control in the form
\beqs
\hat a(t,x,\mu) & := &  a^*(\partial_\mu v(t,\mu)(x),\partial_x\partial_\mu v(t,\mu)(x)) \\
&=& -  \Big[ x- \bar\mu  -   \frac{1}{2 K(t)}  \Big]\Sigma^{-1} \theta^*,
\enqs
and we then has to consider the solution $X^{\P^b}$ to the McKean-Vlasov SDE \reff{SDEMK} under any (equivalent) prior probability measure $\P^b$, $b$ $\in$ $\Vc_\Theta$, governed by
\beqs
dX_t^{\P^b} &=&  - \Big[  X_t^{\P^b} - \E_b[X_t^{\P^b}] -    \frac{1}{2 K(t)} \Big] (\theta^*)\trans \Sigma^{-1}
[ b_t dt + \sigma dW_t^{b}], \;\; 0 \leq t \leq T,  \; \P^b-\mbox{p.s.}.  
\enqs
By taking expectation under $\P^b$, we get
\beqs
\E_b[X_t^{\P^b}]  &=&  x_0 + \int_0^t  \frac{1}{2 K(s)} (\theta^*)\trans \Sigma^{-1} \E_b[b_s] ds, 
\enqs
and see that, in contrast with the case of covariance matrix uncertainty, this expectation depends on the prior probability measure $\P^b$. 
Consequently, the 
family of processes $\{\hat a(t,X_t^{\P^b},\P^b_{_{X_t^{\P^b}}}), 0\leq t \leq T, \Sigma\in\Vc_\Theta\}$ cannot be aggregated into a universal process $\alpha^*$, which would allow us to conclude that $\alpha^*$ is an optimal strategy.  The main issue in the mean-variance framework, compared to classical (robust) expected utility maximization where the optimal strategy depends  in feedback form  only on the wealth state process, arises from the feedback form dependence  of the optimal wealth process not only upon the wealth process, but also on the expected wealth process, which depends  on the prior probability measure when considering drift uncertainty.  The robust mean-variance and Markowitz problem is then a challenging problem that 
could not  be directly  tackled by our approach and that  we postpone for future research. 
}
\ep
\end{Remark}

\subsection{Example 1: uncertain volatility} \label{secvoluncertain}

We consider the uncertain volatility model in the multivariate case with zero correlation as presented in Example 1:   
$\Theta$ $=$ $\Prod_{i=1}^d[\underline{\sigma_i^2},\bar\sigma_i^2]$ with 
$0<\underline{\sigma_i}$ $\leq$ $\bar\sigma_i$ $<$ $\infty$, $i$ $=$ $1,\ldots,d$, and 
\beqs
\gamma(\theta) &=& 
\left(
\begin{array}{ccc}
\sigma_1^2 & \ldots & 0 \\
\vdots & \ddots & \vdots \\
0 & \ldots & \sigma_d^2
\end{array}
\right), \;\;\; \mbox{ for } \; \theta = (\sigma_1^2,\ldots,\sigma_d^2). 
\enqs
In this case, the risk premium function  is simply given by 
\beqs
R(\theta) \; := \; b\trans\gamma(\theta)^{-1}b &=& \sum_{i=1}^d \frac{b_i^2}{\sigma_i^2},  \;\;\; \mbox{ for } \; \theta = (\sigma_1^2,\ldots,\sigma_d^2),
\enqs
and its is clear that the worst-case scenario corresponds to the covariance matrix $\Sigma^*$ $=$ $\bar\Sigma$ $:=$ $\gamma(\bar\theta)$  with 
$\bar\theta$ $=$  $(\bar\sigma_1^2,\ldots,\bar\sigma_d^2)$, i.e. for the highest marginal volatilities.   

From Theorem \ref{theomain2},  we obtain an explicit optimal portfolio strategy for the robust mean-variance  problem under uncertain volatility:
\beqs
\alpha_t^* &=&  \Big[ x_0 + \frac{1}{2\lambda} \exp\big(   \bar R  \; T  \big) - X_t^* \Big]\bar\Sigma^{-1} b, \;\;\; 0 \leq t \leq T, \;\; \Pc^\Theta q.s., \\
\mbox{ with } \;\; \bar R &:= & R(\bar\theta) \; = \; b\trans\bar\Sigma^{-1} b.
\enqs
This corresponds to the optimal mean-variance portfolio strategy in a multidimensional Black-Scholes model with uncorrelated assets of drift $b$ and  covariance matrix   $\bar\Sigma$, as derived in \cite{zholi00} and \cite{fisliv15}.  The financial interpretation is natural: the worst-case scenario corresponds to the highest variance $\bar\Sigma$, and 
the risk-averse investor makes her/his portfolio decision by referring  to this case.

\subsection{Example 2: ambiguous correlation} \label{seccorrel}

We consider the model for a two-risky assets model with ambiguous correlation, i.e.  $\Theta$ $=$ $[\underline{\varrho},\bar\varrho]$ $\subset$ 
$(-1,1)$, and 
\beqs
\gamma(\theta) &=& \Big( 
\begin{array}{cc}
\sigma_1^2   & \sigma_1\sigma_2 \theta \\
\sigma_1\sigma_2\theta & \sigma_2^2 
\end{array}
\Big), \;\;\; \theta \in \Theta, 
\enqs
for some known positive constants $\sigma_1$ $>$ $0$ and $\sigma_2$ $>$ $0$.  

In this case, the risk premium function is given by 
\beq \label{Rcorrel}
R(\theta) \; := \; b\trans\gamma(\theta)^{-1}b &=& \frac{1}{1-\theta^2} \big( \beta_1^2 + \beta_2^2 - 2 \beta_1 \beta_2 \theta \big),
\enq
where we denote by $\beta_i$ $=$ $\frac{b_i}{\sigma_i}$, $i$ $=$ $1,2$,  the instantaneous Sharpe ratio of  each  risky asset. When 
the asset $S^i$ is a stock, its sharpe ratio is usually positive (otherwise it  would perform less than the riskless bond). We may also want to consider the case when $\beta_i$ is nonpositive, which would correspond typically to the case when the asset $S^i$ is  a spread between two stocks. 
In the sequel, we shall assume w.l.o.g. that $(\beta_1,\beta_2)$ $\neq$ $(0,0)$ (in this trivial case, the optimal portfolio strategy is clearly to 
never trade, i.e. $\alpha^*$ $\equiv$ $0$), and we set: 
\beq \label{rho+-}
\varrho_0^+ \; := \; \frac{\min(|\beta_1|,|\beta_2|)}{\max(|\beta_1|,|\beta_2|)} \; \in \; [0,1],  & & 
\varrho_0^- \; := \; - \varrho_0^+. 
\enq
Let us also  introduce the extremal covariance matrices
\beqs
\bar\Sigma \; := \; \gamma(\bar\varrho)  \; = \;  
\Big( 
\begin{array}{cc}
\sigma_1^2   & \sigma_1\sigma_2 \bar\varrho \\
\sigma_1\sigma_2 \bar\varrho & \sigma_2^2 
\end{array}
\Big), & &  \underline{\Sigma} \; := \; \gamma(\underline{\varrho})   \; = \;  
\Big( 
\begin{array}{cc}
\sigma_1^2   & \sigma_1\sigma_2 \underline{\varrho} \\
\sigma_1\sigma_2 \underline{\varrho} & \sigma_2^2 
\end{array}
\Big),
\enqs
and their corresponding variance risk ratios:
\beqs
\bar\Sigma^{-1}b \;  = \; \frac{1}{1-\bar\varrho^2}
\left( 
\begin{array}{c}
\frac{b_1}{\sigma_1^2} - \frac{b_2 \bar\varrho}{\sigma_1\sigma_2}   \\
\frac{b_2}{\sigma_2^2} - \frac{b_1 \bar\varrho}{\sigma_1\sigma_2} 
\end{array}
\right)
\; =: \; 
\Big( 
\begin{array}{c}
\bar\kappa_1    \\
\bar\kappa_2
\end{array}
\Big), & & 
\underline{\Sigma}^{-1}b \; = \;  \frac{1}{1-\underline{\varrho}^2}  
\left( 
\begin{array}{c}
\frac{b_1}{\sigma_1^2} - \frac{b_2 \underline{\varrho}}{\sigma_1\sigma_2}   \\
\frac{b_2}{\sigma_2^2} - \frac{b_1 \underline{\varrho}}{\sigma_1\sigma_2} 
\end{array}
\right)
\; =: \; 
\Big( 
\begin{array}{c}
\underline{\kappa}_1    \\
\underline{\kappa}_2
\end{array}
\Big).
\enqs

The following result provides the explicit determination of the correlation $\theta^*$ achieving  the minimal risk premium.

\begin{Lemma} \label{minrisk}
We distinguish two cases depending on the sign of $\beta_1\beta_2$. 

\noindent {\bf I.} For $\beta_1\beta_2$ $>$ $0$,  we  have: 
\begin{itemize}
\item[{\bf 1.}] if $\bar\varrho$ $<$ $\varrho_0^+$, then $\theta^*$ $=$ $\bar\varrho$. 
Moreover, $\bar\kappa_1\bar\kappa_2$  $>$ $0$ and $\underline{\kappa}_1\underline{\kappa}_2$ $>$ $0$.  
\item[{\bf 2.}] if $\underline{\varrho}$ $>$ $\varrho_0^+$, then  $\theta^*$ $=$ $\underline{\varrho}$.
Moreover, $\underline{\kappa}_1\underline{\kappa}_2$ $<$ $0$ and  $\bar\kappa_1\bar\kappa_2$ $<$ $0$. 
\item[{\bf 3.}] if $\varrho_0^+$ $\in$ $\Theta$ $=$ $[\underline{\varrho},\bar\varrho]$, then $\theta^*$ $=$ $\varrho_0^+$.  
Moreover, $\underline{\kappa}_1\underline{\kappa}_2$ $\geq$ $0$ and  $\bar\kappa_1\bar\kappa_2$ $\leq$ $0$. 
\end{itemize}
\noindent {\bf I'.} For $\beta_1\beta_2$ $\leq$ $0$, we have:
\begin{itemize}
\item[{\bf 1'.}] if $\bar\varrho$ $<$ $\varrho_0^-$, then $\theta^*$ $=$ $\bar\varrho$. 
Moreover, $\bar\kappa_1\bar\kappa_2$  $>$ $0$ and $\underline{\kappa}_1\underline{\kappa}_2$ $>$ $0$.  
\item[{\bf 2'.}] if $\underline{\varrho}$ $>$ $\varrho_0^-$, then $\theta^*$ $=$ $\underline{\varrho}$.
Moreover, $\underline{\kappa}_1\underline{\kappa}_2$ $<$ $0$ and  $\bar\kappa_1\bar\kappa_2$ $<$ $0$. 
\item[{\bf 3'.}]  if $\varrho_0^-$ $\in$ $\Theta$ $=$ $[\underline{\varrho},\bar\varrho]$, then $\theta^*$ $=$ $\varrho_0^-$.  
Moreover, $\underline{\kappa}_1\underline{\kappa}_2$ $\geq$ $0$ and  $\bar\kappa_1\bar\kappa_2$ $\leq$ $0$. 
\end{itemize}
\end{Lemma}
{\bf Proof.}  The risk premium function $R$ is differentiable on $\Theta$ $=$ $[\underline{\varrho},\bar\varrho]$, with a derivative given by: 
\beqs
R'(\theta) \; = \; - \frac{2}{(1-\theta^2)^2}  f(\theta), & \mbox{ with } &  f(\theta) \; = \; \beta_1\beta_2 (1+\theta^2) - (\beta_1^2 + \beta_2^2)\theta. 
\enqs
For any $\theta$ $\in$ $\Theta$,  let us also denote by $\kappa_1(\theta)$, $\kappa_2(\theta)$ 
the components of the variance risk ratio $\Sigma(\theta)^{-1}b$, i.e. 
\beqs
\kappa_1(\theta) \; = \; \frac{1}{1-\theta^2}\Big( \frac{b_1}{\sigma_1^2} - \frac{b_2\theta}{\sigma_1\sigma_2} \Big), & &
\kappa_2(\theta) \; = \; \frac{1}{1-\theta^2}\Big( \frac{b_2}{\sigma_2^2} - \frac{b_1\theta}{\sigma_1\sigma_2} \Big),
\enqs
so that $\bar\kappa_i$ $=$ $\kappa_i(\bar\varrho)$, and $\underline{\kappa}_i$ $=$ $\kappa_i(\underline{\varrho})$, $i$ $=$ $1,2$, and notice that
\beq \label{relkappaf}
\kappa_1(\theta)\kappa_2(\theta) & = &  \frac{1}{\sigma_1\sigma_2 (1-\theta^2)^2}  f(\theta).
\enq
\noindent {\bf I.} We first consider the case when $\beta_1\beta_2$ $>$ $0$. In this case, 
the function $f$ is a strictly convex parabolic function attaining its infimum on $\R$ at $\bar\theta$ $=$ 
$\frac{\beta_1^2+\beta_2^2}{2\beta_1\beta_2}$ $\geq$ $1$, which implies that $f$ is strictly decreasing on $(-\infty,\bar\theta]$ hence on $\Theta$. 
Since  $f(0)$ $=$ $\beta_1\beta_2$ $>$ $0$ and  $f(1)$ $=$ $-(\beta_1-\beta_2)^2$ $\leq$ $0$, there exists a unique 
$\varrho_0^+$ $\in$ $(0,1]$ s.t. $f(\varrho_0^+)$ $=$ $0$, which is exactly given by the expression in \reff{rho+-}. 
We are then led to distinguish the following cases:

\vspace{1mm}

\noindent {\bf 1.} $\bar\varrho$ $<$ $\varrho_0^+$.

\noindent In this case, recalling that $f$ is strictly decreasing on $\Theta$ $=$ $[\underline{\varrho},\bar\varrho]$, 
we see that  for all $\theta$  $\in$  $\Theta$, $f(\theta)$ $>$ $f(\varrho_0^+)$ $=$ $0$, i.e.  $R'(\theta)$ $<$ $0$ on $\Theta$, i.e. 
$R$ is strictly decreasing on $\Theta$, and thus: $\theta^*$ $=$ ${\rm arg}\min_{\theta\in\Theta} R(\theta)$ $=$  $\bar\varrho$.  
Moreover, by \reff{relkappaf}, we have $\kappa_1(\theta)\kappa_2(\theta)$ $>$ $0$ for all $\theta$ $\in$ $\Theta$, 
and thus:   $\underline{\kappa}_1\underline{\kappa}_2$ $>$ $0$ and 
$\bar\kappa_1\bar\kappa_2$ $>$ $0$. 
 
\noindent {\bf 2.}  $\underline{\varrho}$ $>$ $\varrho_0^+$.  In this case, 
$f(\bar\varrho)$ $\leq$ $f(\underline{\varrho})$ $<$ $f(\varrho_0^+)$ $=$ $0$, and thus for all $\theta$  $\in$  $\Theta$, $f(\theta)$ $<$ $0$, 
$\kappa_1(\theta)\kappa_2(\theta)$ $<$ $0$,  $R'(\theta)$ $>$ $0$, i.e. 
$R$ is strictly increasing on $\Theta$. This implies that $\theta^*$ $=$ ${\rm arg}\min_{\theta\in\Theta} R(\theta)$ $=$ $\underline{\varrho}$, and also  
$\underline{\kappa}_1\underline{\kappa}_2$ $<$ $0$, $\bar\kappa_1\bar\kappa_2$ $<$ $0$.

\noindent  {\bf 3.} $\underline{\varrho}$ $\leq$ $\varrho_0^+$ $\leq$ $\bar\varrho$, i.e. $\varrho_0^+$ $\in$ $\Theta$.  Notice that in this case, 
$\varrho_0^+$ is strictly smaller than $1$ (recall that $\bar\varrho$ $<$ $1$), and thus $\beta_1$ $\neq$ $\beta_2$. 
Again, since $f$ is decreasing, we have  $f(\theta)$ $\geq$ $f(\varrho_0^+)$ $=$ $0$ for $\theta$ $\in$ $[\underline{\varrho},\varrho_0^+]$, and  
$f(\theta)$ $\leq$ $f(\varrho_0^+)$ $=$ $0$ for $\theta$ $\in$ $[\varrho_0^+,\bar\varrho]$.  Therefore, $\kappa_1(\theta)\kappa_2(\theta)$ $\geq$ $0$, 
$R'(\theta)$ $\leq$ $0$ for $\theta$ $\in$ $[\underline{\varrho},\varrho_0^+]$, i.e. $R$ is decreasing on $[\underline{\varrho},\varrho_0^+]$, and 
$\kappa_1(\theta)\kappa_2(\theta)$ $\leq$ $0$, $R'(\theta)$ $\geq$ $0$ for $\theta$ $\in$ $[\varrho_0^+,\bar\varrho]$, i.e. 
$R$ is increasing on $[\varrho_0^+,\bar\varrho]$. Therefore,  $\theta^*$ $=$ ${\rm arg}\min_{\theta\in\Theta} R(\theta)$ $=$ $\varrho_0^+$, and we also have
$\underline{\kappa}_1\underline{\kappa}_2$ $\geq$ $0$ and  $\bar\kappa_1\bar\kappa_2$ $\leq$ $0$. 

\vspace{1mm}

\noindent {\bf I'.} We finally  consider the case when $\beta_1\beta_2$ $\leq$ $0$.  When $\beta_1\beta_2$ $<$ $0$,  
the function $f$ is a strictly concave parabolic function attaining its infimum on $\R$ at $\bar\theta$ $=$ 
$\frac{\beta_1^2+\beta_2^2}{2\beta_1\beta_2}$ $\leq$ $-1$, and when $\beta_1\beta_2$ $=$ $0$,  $f$ is a linear function with strictly negative slope.  
In any case, the function  $f$ is strictly decreasing on $[\bar\theta,\infty)$ hence on $\Theta$.  
Since  $f(0)$ $=$ $\beta_1\beta_2$ $\leq$ $0$, $f(-1)$ $=$ $(\beta_1+ \beta_2)^2$ $\geq$ $0$, there exists a unique 
$\varrho_0^-$ $\in$ $[-1,0]$ s.t. $f(\varrho_0^-)$ $=$ $0$, which is exactly given by the expression in \reff{rho+-}, i.e. $\varrho_0^-$ $=$ $-\varrho_0^+$. Then, 
by distinguishing the cases when $\varrho_0^-$ $>$ $\underline{\varrho}$, $\varrho_0^-$ $<$ $\underline{\varrho}$  and $\varrho_0^-$ $\in$ $\Theta$, and 
proceeding by the same arguments as in Case {\bf I}, we obtain the results described in {\bf 1'}, {\bf 2'} and {\bf 3'}. 
\ep

\vspace{3mm}

By applying Theorem \ref{theomain2}, we can now provide an  explicit description of the optimal strategy
under ambiguous correlation.

\begin{Theorem} \label{procor}
The solution to  problem \reff{robustMV} is explicitly described through  the following cases\footnote{By misuse of notation, we write indifferently $a$ $=$ 
$(a_1,a_2)$ or  $a$ $=$ 
$\Big( \begin{array}{c} a_1 \\ a_2 \end{array} \Big)$ for an element in $\R^2$.  
} :
\begin{itemize}
\item[{\bf I.}] If $\beta_1\beta_2$ $>$ $0$, and 
\begin{itemize}
\item[{\bf 1.}]  $\bar\varrho$ $<$ $\varrho_0^+$,  then an  optimal portfolio strategy is explicitly given by
\beqs
\alpha_t^* &=& \Big[ x_0 + \frac{1}{2\lambda} \exp\big(  \bar R  T  \big) - X_t^* \Big] \bar\Sigma^{-1} b ,  \;\;\; 0 \leq t \leq T, \;\; \Pc^\Theta q.s., 
\enqs
with $\bar R$ $=$ $b\trans\bar\Sigma^{-1} b$, and the optimal cost is 
\beqs
V_0 &=& - \frac{1}{4\lambda}  \big[ \exp\big(   \bar R  T  \big)  -1 \big]  - x_0. 
\enqs
\item[{\bf 2.}]  $\underline{\varrho}$ $>$ $\varrho_0^+$,  then an  optimal portfolio strategy is explicitly given by 
\beqs
\alpha_t^* &=&  \Big[ x_0 + \frac{1}{2\lambda} \exp\big(  \underline{R} T  \big) - X_t^* \Big] \underline{\Sigma}^{-1} b, 
\;\;\; 0 \leq t \leq T, \;\; \Pc^\Theta q.s.,  
\enqs
with $\underline{R}$ $=$ $b\trans\underline{\Sigma}^{-1} b$, and the optimal cost is
\beqs
V_0 &=& -\frac{1}{4\lambda}  \big[ \exp\big(  \underline{R}  T  \big)  -1 \big]  - x_0.
\enqs
\item[{\bf 3.}] $\underline{\varrho}$ $\leq$ $\varrho_0^+$ $\leq$ $\bar\varrho$, then an  optimal portfolio strategy is explicitly given by  
\beqs
\alpha_t^* &=& 
\left\{
\begin{array}{cl}
\left(
\begin{array}{c}
\Big[ x_0 + \frac{1}{2\lambda} \exp\big(  \beta_1^2  T  \big) - X_t^* \Big] \frac{b_1}{\sigma_1^2}  \\
0
\end{array}
\right),  \;\; 0 \leq t \leq T, \;\; \Pc^\Theta q.s., & \mbox{when }  \beta_1^2 > \beta_2^2, \\
\\
\left(
\begin{array}{c}
0 \\
\Big[ x_0 + \frac{1}{2\lambda} \exp\big(   \beta_2^2  T  \big) - X_t^* \Big] \frac{b_2}{\sigma_2^2}
\end{array}
\right),  \;\; 0 \leq t \leq T, \;\; \Pc^\Theta q.s., & \mbox{when }  \beta_2^2 > \beta_1^2. 
\end{array}
\right.
\enqs
and the optimal cost
\beqs
V_0 &=& -\frac{1}{4\lambda}  \big[ \exp\big(  \max(\beta_1^2,\beta_2^2) T  \big)  -1 \big]  - x_0.
\enqs
\end{itemize}
\item[{\bf I'.}] If $\beta_1\beta_2$ $\leq$ $0$, and 
\begin{itemize}
\item[{\bf 1'.}]   $\bar\varrho$ $<$ $\varrho_0^-$,  then an  optimal portfolio strategy is explicitly given by 
\beqs
\alpha_t^* &=&  \Big[ x_0 + \frac{1}{2\lambda} \exp\big(   \bar R T  \big) - X_t^* \Big] \bar\Sigma^{-1} b, \;\;\; 0 \leq t \leq T, \;\; \Pc^\Theta q.s., 
\enqs
and the optimal cost is
\beqs
V_0 &=& -\frac{1}{4\lambda}  \big[ \exp\big(  \bar R T  \big)  -1 \big]  - x_0.
\enqs
Moreover, $\bar\kappa_1\bar\kappa_2$  $>$ $0$ and $\underline{\kappa}_1\underline{\kappa}_2$ $>$ $0$.  
\item[{\bf 2'.}]   $\underline{\varrho}$ $>$ $\varrho_0^-$,  then an  optimal portfolio strategy is explicitly given by  
\beqs
\alpha_t^* &=&  \Big[ x_0 + \frac{1}{2\lambda} \exp\big(   \underline{R}  T  \big) - X_t^* \Big] \underline{\Sigma}^{-1} b, 
\;\;\; 0 \leq t \leq T, \;\; \Pc^\Theta q.s.,  
\enqs
and the optimal cost
\beqs
V_0 &=& - \frac{1}{4\lambda}  \big[ \exp\big(  \underline{R}  T  \big)  -1 \big]  - x_0.
\enqs
Moreover, $\underline{\kappa}_1\underline{\kappa}_2$ $<$ $0$ and  $\bar\kappa_1\bar\kappa_2$ $<$ $0$. 
\item[{\bf 3'.}]   $\underline{\varrho}$ $\leq$ $\varrho_0^-$ $\leq$ $\bar\varrho$, then an  optimal portfolio strategy is explicitly given by  
\beqs
\alpha_t^* &=& 
\left\{
\begin{array}{cl}
\left(
\begin{array}{c}
\Big[ x_0 + \frac{1}{2\lambda} \exp\big(  \beta_1^2  T  \big) - X_t^* \Big] \frac{b_1}{\sigma_1^2}  \\
0
\end{array}
\right),  \;\; 0 \leq t \leq T, \;\; \Pc^\Theta q.s., & \mbox{when }  \beta_1^2 > \beta_2^2, \\
\\
\left(
\begin{array}{c}
0 \\
\Big[ x_0 + \frac{1}{2\lambda} \exp\big(   \beta_2^2  T  \big) - X_t^* \Big] \frac{b_2}{\sigma_2^2}
\end{array}
\right),  \;\; 0 \leq t \leq T, \;\; \Pc^\Theta q.s., & \mbox{when }  \beta_2^2 > \beta_1^2. 
\end{array}
\right.
\enqs
and the optimal cost
\beqs
V_0 &=& -\frac{1}{4\lambda}  \big[ \exp\big(  \max(\beta_1^2,\beta_2^2) T  \big)  -1 \big]  - x_0.
\enqs
\end{itemize}
Moreover, $\underline{\kappa}_1\underline{\kappa}_2$ $\geq$ $0$ and  $\bar\kappa_1\bar\kappa_2$ $\leq$ $0$.   
\end{itemize}
\end{Theorem} 
{\bf Proof.} In view of the formulae  \reff{explialpha*gen} and \reff{Vopti}  of the optimal portfolio strategy and optimal cost in Theorem \ref{theomain2}, we only has to compute the minimal risk premium $R^*$ $=$ $R(\theta^*)$, and the vector $(\Sigma^*)^{-1}b$ with $\Sigma^*$ $=$ $\gamma(\theta^*)$, and $\theta^*$ explicitly given in Lemma \ref{minrisk}. 
We only consider the case {\bf I} when $\beta_1\beta_2$ $>$ $0$ since the other case {\bf I'}  is dealt with similarly.  The subcases {\bf 1} and {\bf 2} are immediate, and we only focus on the third case {\bf 3} when  $\varrho_0^+$ $\in$ $[\underline{\varrho},\bar\varrho]$. In this case $\theta^*$ $=$ $\varrho_0^+$, and a simple computation from  the expressions of $R$ in  \reff{Rcorrel} and 
$\varrho_0^+$ in \reff{rho+-}  gives:  $R^*$ $=$ $R(\varrho_0^+)$ $=$ $\max(\beta_1^2,\beta_2^2)$. Moreover, a straightforward calculation shows that
\beqs
(\Sigma^*)^{-1}b \; = \; \gamma(\varrho_0^+)^{-1} b &=& 
\left\{
\begin{array}{cl}
\left(
\begin{array}{c}
\frac{b_1}{\sigma_1^2}  \\
0
\end{array}
\right),  & \mbox{when }  \beta_1^2 > \beta_2^2, \\
\\
\left(
\begin{array}{c}
0 \\
\frac{b_2}{\sigma_2^2}
\end{array}
\right),  & \mbox{when }  \beta_2^2 > \beta_1^2, 
\end{array}
\right. 
\enqs
which leads to the expression of the optimal portfolio strategy  in the assertion of the Theorem. 
\ep

\vspace{2mm}

\begin{Remark}
{\rm (Financial interpretation) 

\noindent 
To fix the idea, we focus on the usual case of two stocks when $\beta_1$ $>$ $0$, $\beta_2$ $>$ $0$. The coefficient $\varrho_0^+$ can be viewed as a measure for the ``proximity"  between the two stocks: a small $\varrho_0^+$ (close to zero) means that one stock  is much better than the other one in the sense that it has a  much larger instantaneous Sharpe ratio,  
while large $\varrho_0^+$ (close to one) means that the two stocks are similar in terms of instantaneous Sharpe ratio. 
 
When $\bar\varrho$ $<$ $\varrho_0^+$, this means that  no stock is ``dominating" the other one, and  it is  optimal to invest in both assets with a directional trading, that is buying or selling simultaneously (recall from Lemma \ref{minrisk}  that in this case 
$\bar\kappa_1\bar\kappa_2$ $>$ $0$), and  the worst-case scenario refers to the highest corre\-lation 
$\bar\rho$ where the diversification effect is minimal.  The optimal strategy corresponds to the optimal mean-variance portfolio strategy in a market with constant covariance matrix $\bar\Sigma$. 

When $\underline{\varrho}$ $>$ $\varrho_0^+$, this means that one asset is clearly dominating the other one, and it is  optimal to invest in both assets with a spread trading, that is buying one and selling another 
(recall that in this case $\underline{\kappa}_1\underline{\kappa}_2$ $<$ $0$), and the worst-case scenario  corresponds to the lowest correlation where the 
profit from the spread trading is minimal.  

When $\underline{\varrho}$ $\leq$ $\varrho_0^+$ $\leq$ $\bar\varrho$, it is optimal to invest in either one of the stocks, but not both, since the directional trading is not optimal for high correlation and the spread trading is not optimal for low correlation. The selection for the risky asset is then naturally  made on the one with the  highest instantaneous Sharpe ratio.  

We notice that a similar interpretation was derived in \cite{fouetal15} for robust portfolio optimization with utility function, but in this cited paper, the authors derived the worst-case scenario by distinguish four cases (see their Theorem 2.2):   
(1) $\bar\kappa_1\bar\kappa_2$ $>$ $0$ and $\underline{\kappa}_1\underline{\kappa}_2$ $\geq$ $0$, 
(2) $\bar\kappa_1\bar\kappa_2$ $\leq$ $0$ and $\underline{\kappa}_1\underline{\kappa}_2$ $<$ $0$, (3) $\bar\kappa_1\bar\kappa_2$ $\leq$ $0$ and $\underline{\kappa}_1\underline{\kappa}_2$ $\geq$ $0$, and (4) $\bar\kappa_1\bar\kappa_2$ $>$ $0$ and $\underline{\kappa}_1\underline{\kappa}_2$ $<$ $0$. 
Compared to \cite{fouetal15}, we push further the calculations and reduce the different cases on  the variance risk  ratios $\bar\kappa_1$ and $\bar\kappa_2$ to an explicit description with three cases in terms of the correlations $\bar\varrho$, $\underline{\varrho}$, and $\varrho_0^+$. 
Actually, as shown in Lemma \ref{minrisk}, our cases {\bf 1}, resp. {\bf 2}, resp.  {\bf 3} in Theorem \ref{procor} 
are equivalent to their cases (1), resp. (2), resp.  (3), and it appears that their last case (4) can never happen. 
Let us also mention  that a similar description with three cases in terms of the correlation  was done  in \cite{liuzhe16} (see their Proposition 2) for  a single-period mean-variance problem under correlation ambiguity.  
}
\ep
\end{Remark}

\section{Robust efficient frontier}

\setcounter{equation}{0}
\setcounter{Assumption}{0} \setcounter{Theorem}{0}
\setcounter{Proposition}{0} \setcounter{Corollary}{0}
\setcounter{Lemma}{0} \setcounter{Definition}{0}
\setcounter{Remark}{0}

Let us denote by $U_0(\vartheta)$ the optimal worst-case expected terminal wealth  given a worst-case variance risk $\vartheta$ $>$ $0$, i.e., 
\beqs
U_0(\vartheta) &=& \sup \Big\{ \Ec(\alpha): \alpha\in\Ac, \Rc(\alpha) \leq  \vartheta \},
\enqs
where we recall the notations  from the robust Markowitz problem \reff{robustMarkowitz}:
\beqs
\Ec(\alpha) \; := \; \inf_{\P^\sigma\in\Pc^\Theta}  \E_\sigma[X_T^\alpha], & & 
\Rc(\alpha) \; := \; \sup_{\P^\sigma\in\Pc^\Theta} {\rm Var_\sigma}(X_T^\alpha).
\enqs
By the linearity of $X^\alpha$ w.r.t.  $\alpha$ lying in the convex set $\Ac$, the convexity (resp. the linearity) 
of $X$ $\in$ $L^2(\Fc_T,\P^\sigma)$ $\mapsto$ ${\rm Var_\sigma}(X)$ (resp. $\E_\sigma[X]$), it is easily seen that the function $U_0$ is concave w.r.t.  
$\vartheta$ $\in$ $(0,\infty)$.

We consider  the general  framework of Section \ref{secexpli} under condition {\bf (IC)}, and  
emphasize the dependence of $V_0$ $=$ $V_0(\lambda)$, and $\alpha^*$ $=$ $\alpha^{*,\lambda}$, for the optimal cost and optimal 
portfolio strategy to the robust mean-variance portfolio selection problem \reff{robustMV} with risk-aversion parameter $\lambda$:
\beqs
V_0(\lambda) &=& \inf_{\alpha\in\Ac} \sup_{\P^\sigma\in\Pc^\Theta} 
\Big( \lambda  {\rm Var_\sigma}(X_T^\alpha) -  \E_\sigma[X_T^\alpha] \Big) \\
&=&  \sup_{\P^\sigma\in\Pc^\Theta} 
\Big( \lambda  {\rm Var_\sigma}(X_T^{\alpha^{*,\lambda}} ) -  \E_\sigma[X_T^{\alpha^{*,\lambda}}] \Big). 
\enqs
From \reff{Vopti} in Theorem \ref{theomain2}, we recall that
\beq \label{V0expli} 
V_0(\lambda) &=& - \frac{1}{4\lambda} \Big[  \exp\big( R(\theta^*) T \big) - 1 \Big]   - x_0,
\enq
where $R(\theta)$ $=$ $b\trans\gamma(\theta)^{-1}b$,  and $\theta^*$ $=$  
${\rm arg}\min_{\theta\in\Theta} R(\theta)$.   Moreover, a crucial observation (see \reff{returnopt})  is that the expected optimal terminal wealth $\E_\sigma[X_T^{\alpha^{*,\lambda}}]$ under any prior probability measure $\P^\sigma$ does not depend actually on $\Sigma$ $\in$  $\Vc_\Theta$, and thus:
\beq \label{Eccrucial}
\Ec(\alpha^{*,\lambda}) &=&  \E_\sigma[X_T^{\alpha^{*,\lambda}}] \; =: \; \bar\rho_T^{*,\lambda},  \;\;\; \forall \Sigma \in \Vc_\Theta, 
\enq
with 
\beq \label{explirho}
\bar\rho_T^{*,\lambda} &=&  x_0 + \frac{1}{2\lambda} \Big[\exp\big(R(\theta^*) T\big) -1 \Big]. 
\enq
By adapting standard arguments from convex optimization theory,  we show the duality relation between the robust mean-variance problem and the robust Markowitz problem, namely:
\begin{equation}  \label{polarUV} 
\begin{array}{ccc}
V_0(\lambda) &=& \inf_{\vartheta > 0} \big[ \lambda \vartheta - U_0(\vartheta) \big], \;\;\; \forall \lambda > 0,   \\
U_0(\vartheta) &=& \inf_{\lambda > 0} \big[   \lambda \vartheta - V_0(\lambda) \big], \;\;\; \forall \vartheta > 0. 
\end{array}
\end{equation}
Indeed,  for  fixed $\vartheta$ $>$ $0$, and for  any $\eps$ $>$ $0$, there exists an $\eps$-optimal control for $U_0(\vartheta)$, that is a control 
$\tilde\alpha^\eps$ $\in$ $\Ac$ s.t. $U_0(\vartheta)$ $\leq$ $\Ec(\tilde\alpha^\eps)$ $+$ $\eps$, and $\Rc(\tilde\alpha^\eps)$ $\leq$ $\vartheta$. It follows that for all $\lambda$ $>$ $0$, 
\beqs
V_0(\lambda) & \leq & \sup_{\P^\sigma\in\Pc^\Theta} 
\Big( \lambda  {\rm Var_\sigma}(X_T^{\tilde\alpha^\eps}) -  \E_\sigma[X_T^{\tilde\alpha^\eps}] \Big) \\
& \leq & \lambda   \sup_{\P^\sigma\in\Pc^\Theta}   {\rm Var_\sigma}(X_T^{\tilde\alpha^\eps}) -  \inf_{\P^\sigma\in\Pc^\Theta} 
\E_\sigma[X_T^{\tilde\alpha^\eps}] \; = \; \lambda  \Rc(\tilde\alpha^\eps)  -  \Ec(\tilde\alpha^\eps) \\
& \leq & \lambda \vartheta - U_0(\vartheta)  + \eps. 
\enqs
Since $\eps$ is arbitrary, and the above relation holds for any fixed $\vartheta$ $>$ $0$, this shows that
\beq \label{ineg1polar}
V_0(\lambda) & \leq & \inf_{\vartheta > 0} \big[ \lambda \vartheta - U_0(\vartheta) \big], \;\;\; \forall \lambda > 0. 
\enq
Conversely, for fixed $\lambda$ $>$ $0$, let us consider the  optimal control $\alpha^{*,\lambda}$ $\in$ $\Ac$  for $V_0(\lambda)$, and set 
$\vartheta_\lambda$ $:=$ $\Rc(\alpha^{*,\lambda})$ which is strictly positive since the terminal wealth 
$X_T^{\alpha^{*,\lambda}}$ is not constant.  Then, by definition of $U_0(\vartheta_\lambda)$,   we have 
$\Ec(\alpha^{*,\lambda})$ $\leq$ $U_0(\vartheta_\lambda)$, and so by \reff{Eccrucial} 
\beq
V_0(\lambda)  &=& \sup_{\P^\sigma\in\Pc^\Theta} 
\Big( \lambda  {\rm Var_\sigma}(X_T^{\alpha^{*,\lambda}} ) -  \E_\sigma[X_T^{\alpha^{*,\lambda}}] \Big)  \; = \;  \lambda \Rc(\alpha^{*,\lambda}) - \Ec(\alpha^{*,\lambda})  \label{V00} \\
& \geq & \lambda \vartheta_\lambda - U_0(\vartheta_\lambda).  \nonumber
\enq
Together with \reff{ineg1polar}, this shows the first duality relation in \reff{polarUV}, i.e., $V_0$ is the Fenchel-Legendre transform of $U_0$, and 
$\vartheta_\lambda$ attains the infimum in this transform:
\beq \label{V01}
V_0(\lambda) & = & \inf_{\vartheta > 0} \big[ \lambda \vartheta - U_0(\vartheta) \big] \; = \; \lambda \vartheta_\lambda - U_0(\vartheta_\lambda).
\enq
By concavity of $U_0$, we deduce (see e.g. \cite{roc70})  
the second duality relation in  \reff{polarUV}, i.e., $U_0$ is the Fenchel-Legendre transform of $V_0$. 

Next, observe from the explicit expression of $V_0$ in \reff{V0expli}, that $V_0$ is a strictly concave $C^1$ function on $(0,\infty)$, with 
$V_0'(0^+)$ $=$ $\infty$, $V_0'(\infty)$ $=$ $0$. Then, for any fixed $\vartheta$ $>$ $0$, there exists a unique $\lambda_\vartheta$ $>$ $0$ that attains the infimum of $\lambda$ $\in$ $(0,\infty)$ $\mapsto$ $\lambda\vartheta-V_0(\lambda)$, characterized by $V_0'(\lambda_\vartheta)$ $=$ $\vartheta$, and explicitly given by 
\beq \label{lambdavartheta}
\lambda_\vartheta &=& \sqrt{\frac{\exp\big(R(\theta^*) T\big) -1 }{4\vartheta}}. 
\enq
Relation \reff{lambdavartheta} gives the explicit link between the variance risk  in the robust Markowitz problem and the Lagrange multiplier  in the robust mean-variance problem. This Lagrange multiplier $\lambda$ 
is then interpreted as a risk-aversion parameter: the larger is $\lambda_\vartheta$, the lower is the variance risk $\vartheta$.  
From the duality relation \reff{polarUV},  we then have:  
\beqs
V_0(\lambda_\vartheta) &=& \lambda_\vartheta\vartheta - U_0(\vartheta) \; = \; \inf_{\vartheta'>0} [ \lambda_\vartheta \vartheta' - U_0(\vartheta') ],
\enqs
which means that $\vartheta$ attains the infimum of $\vartheta'$ $\in$ $(0,\infty)$ $\mapsto$ $\lambda_\vartheta \vartheta' - U_0(\vartheta')$. 
Since $V_0$ is strictly concave, its Fenchel-Legendre transform $U_0$ is also strictly concave (see e.g. \cite{roc70}), and thus this infimum is unique. 
Recalling \reff{V01}, this shows that $\vartheta$ $=$ $\vartheta_{\lambda_\vartheta}$ $=$ $\Rc(\alpha^{*,\lambda_\vartheta})$. Together with \reff{V00}, we then obtain:
\beqs
U_0(\vartheta) &=& \lambda_\vartheta\vartheta  -  V_0(\lambda_\vartheta) \\
&=&  \lambda_\vartheta \Rc(\alpha^{*,\lambda_\vartheta}) 
- \big[  \lambda_\vartheta \Rc(\alpha^{*,\lambda_\vartheta}) - \Ec(\alpha^{*,\lambda_\vartheta}) \big] \;  = \;  \Ec(\alpha^{*,\lambda_\vartheta}),
\enqs
which proves that $\hat\alpha^\vartheta$ $=$ $\alpha^{*,\lambda_\vartheta}$ is a solution to the robust Markowitz problem $U_0(\vartheta)$, 
i.e., a robust efficient portfolio strategy given a worst-case variance risk $\vartheta$ $>$ $0$.  From \reff{Eccrucial}, \reff{explirho} and \reff{lambdavartheta}, we get the explicit form of the robust efficient frontier:
\beq 
U_0(\vartheta) \; = \; \Ec(\hat\alpha^\vartheta) & = & \bar\rho_T^{*,\lambda_\vartheta} \nonumber \\
&=& x_0 + \sqrt{\vartheta} \sqrt{ \exp\big(R(\theta^*) T\big) -1}, \;\;\; \vartheta > 0  \label{refficient} \\
&=& x_0 + \sqrt{\Rc(\hat\alpha^\vartheta)} \sqrt{ \exp\big(R(\theta^*) T\big) -1}.  \nonumber 
\enq

To summarize the above discussion, we have the following result: 

\begin{Theorem}
Under {\bf (IC)}, the efficient frontier of the robust Markowitz problem \reff{robustMarkowitz} is explicitly given by the relation \reff{refficient}.
\end{Theorem}

The relation \reff{refficient} determines explicitly the tradeoff between the worst-case mean (return) and worst-case variance (risk), and can be inverted: 
given an expected  return level $m$ $>$ $x_0$,  the risk that the robust investor can take is: 
\beqs
\hat\vartheta(m) \; = \; U_0^{-1}(m) &=&  \frac{(m-x_0)^2}{ \exp\big(R(\theta^*) T\big) -1   }, \;\;\; m > x_0.    
\enqs
Notice that the robust efficient frontier \reff{refficient} involves a square-root shape as in the classical efficient frontier in Markowitz problem, see e.g. 
\cite{zholi00}.

Let us consider the Sharpe ratio for a portfolio strategy $\alpha$ $\in$ $\Ac$, defined by
\beqs
\Sc(\alpha) &=& \frac{ \E[X_T^\alpha] -x_0}{\sqrt{{\rm Var}(X_T^\alpha)}}, 
\enqs 
that is the excess of the expected return per unit of the standard deviation, evaluated  under the true historical probability measure.  By definition of the 
robust Markowitz problem, and from the relation \reff{refficient}, we have a lower bound for the Sharpe ratio of any robust efficient portfolio strategy 
$\hat\alpha^\vartheta$:
\beqs \label{lowersharpe}
\Sc(\hat\alpha^\vartheta) & \geq & \frac{ \Ec(\hat\alpha^\vartheta) - x_0}{\sqrt{\Rc(\hat\alpha^\vartheta)}} \; = \; \sqrt{ \exp\big(R(\theta^*) T\big) -1} \; =: \; 
\underline{\Sc}.
\enqs
In other words, a robust investor can achieve a Sharpe ratio at least greater than $\underline{\Sc}$ $>$ $0$, and this lower bound is robust to any model misspecification  on the covariance matrix.

\section{Robust Sharpe ratio vs model misspecification}

\setcounter{equation}{0}
\setcounter{Assumption}{0} \setcounter{Theorem}{0}
\setcounter{Proposition}{0} \setcounter{Corollary}{0}
\setcounter{Lemma}{0} \setcounter{Definition}{0}
\setcounter{Remark}{0}

In this section, we illustrate through two examples how robust mean-variance portfolio strategies may help to protect the investor from model misspecification, and can sometimes increase the Sharpe ratio for a specific choice of parameters.


\subsection{A Heston-type  stochastic volatility model}
 
We consider a market with one risky asset, and assume that the true dynamics of the stock price is given by a Heston-type stochastic volatility model
\begin{equation} \label{Heston}
\left\{ \begin{array}{ccl}
dS_t &=& S_t (b dt + \sigma_t dW_t) \\
d\sigma_t^2 &=& \kappa( \sigma_\infty^2 - \sigma_t^2) dt + \eta \sqrt{ (\sigma_t^2 - \underline{\sigma}^2) (\bar\sigma^2 -\sigma_t^2) } d\tilde W_t
\end{array}
\right.
\end{equation}
where $W,\tilde W$ are two  Brownian motions under the real probability measure $\P$,  with negative correlation $\varrho$ representing the leverage effect, 
$\kappa$ $>$ $0$, $\sigma_\infty$ $\in$ $[\underline{\sigma},\bar\sigma]$, $0$ $<$ $\underline{\sigma}$ $\leq$ $\bar\sigma$ $<$ $\infty$. Compared to the original Heston stochastic volatility model where the variance  $\sigma_t^2$ follows a Cox-Ingersoll-Ross process, and is thus valued in $(0,\infty)$, we consider here a variation 
where the variance follows a Wright-Fisher dynamics, and is bounded, valued in $[\underline{\sigma}^2,\bar\sigma^2]$.   

We now consider a simple  investor who  knows the drift $b$ but specifies incorrectly the volatility 
by considering that it is equal to a constant $\tilde\sigma_0$.  In other words, she/he believes that the stock price is governed by a Black-Scholes model of parameters $(b,\tilde\sigma_0)$. Therefore,  from the result in \cite{zholi00} or as a particular case of our paragraph \ref{secvoluncertain}  when $\Theta$ is reduced to the singleton $\{\tilde\sigma_0^2\}$, the optimal mean-variance portfolio strategy of this  
``misspecified" investor with risk-aversion parameter $\lambda$ $>$ $0$, and initial capital $x_0$ is given by:
\beq
\tilde\alpha_t 
&=& \frac{b}{\tilde\sigma_0^2}  \Big[ x_0 + \frac{1}{2\lambda} \exp\big(   \frac{b^2}{\tilde\sigma_0^2} T  \big) - \tilde X_t \Big],  \;\;\; 0 \leq t \leq T, \label{tildealpha}
\enq
where $\tilde X_t$ is the wealth process with  feedback strategy $\tilde\alpha$.  
Notice that the evolution of the wealth process $\tilde X$ under the real probability measure 
$\P$  is
\beqs
d\tilde X_t &=& \tilde\alpha_t \frac{dS_t}{S_t} \; = \; \tilde \alpha_t b dt + \tilde \alpha_t \sigma_t dW_t, 
\enqs
which implies that  its expected return under $\P$  is governed by 
\beqs
d \E[\tilde X_t] &=& b \E[\tilde \alpha_t] dt \; = \; \frac{b^2}{\tilde\sigma_0^2}  \Big[ x_0 + \frac{1}{2\lambda} \exp\big(   \frac{b^2}{\tilde\sigma_0^2} T  \big) - \E[\tilde X_t] \Big] dt.
\enqs
where we used \reff{tildealpha}. Therefore, the excess expected return under $\P$ is explicitly given by:
 \beqs
 \E[\tilde X_T] - x_0 &=& 
 \frac{1}{2\lambda} \Big[ \exp\big(   \frac{b^2}{\tilde\sigma_0^2} T  \big)  - 1\Big]. 
 \enqs
The variance risk  of $\tilde X_T$ under $\P$ is not explicit, but can be approximated by $N$ Monte-Carlo simulations $(\tilde X^i)_{i=1,\ldots,N}$  of $\tilde X$ under $\P$ via:
\beqs
{\rm Var}(\tilde X_T) & \simeq & \frac{1}{N-1} \sum_{i=1}^N \big( \tilde X_T^i - \E[\tilde X_T] \big)^2.
\enqs  
We  can then compute the  Sharpe ratio $\Sc(\tilde\alpha)$ $=$ $\frac{ \E[\tilde X_T] - x_0}{\sqrt{ {\rm Var}(\tilde X_T)}}$  for  the ``misspecified" investor. 

The model parameters used in the simulations for  the bounded Heston stochastic vola\-tility  model \reff{Heston}  are given in Table \ref{tableparam}. 
We used the simulation method as in \cite{deedel98} for dealing with the discretization of the CIR process for the volatility, which means that when a volatility trajectory breachs the bounds 
$\underline{\sigma}$ or $\bar\sigma$, we project its value according to its closest neighbor on $[\underline{\sigma},\bar\sigma]$. 

We fix an investment horizon $T$ $=$ $1$ year, a risk-aversion parameter $\lambda$ $=$ $5$, and use $N$ $=$ $500 000$  
simulations for each set of parameters. 

On the other hand, let us consider a robust investor with risk-aversion parameter $\lambda$, initial capital $x_0$, who knows only the bounds 
$\underline{\sigma}$, $\bar\sigma$ of the volatility,  and  then follows a robust efficient portfolio strategy $\alpha^*$ $=$ $\alpha^{*,\lambda}$ given by
\beqs
\alpha_t^* &=& \frac{b}{\bar\sigma^2}  \Big[ x_0 + \frac{1}{2\lambda} \exp\big(   \frac{b^2}{\bar\sigma^2} T  \big) - X_t^* \Big],  \;\;\; 0 \leq t \leq T. 
\enqs 
Her/his excess expected return under $\P$ is then explicitly given by
 \beqs
 \E[X_T^*] - x_0 &=&  \frac{1}{2\lambda} \Big[ \exp\big(   \frac{b^2}{\bar\sigma^2} T  \big)  - 1\Big]. 
 \enqs
The variance risk of $X_T^*$ under $\P$  is  approximated by Monte-Carlo simulations of  $X^*$ under $\P$, and we  then  compute the  Sharpe ratio 
$\Sc(\alpha^*)$ $=$ $\frac{ \E[X_T^*] - x_0}{\sqrt{ {\rm Var}(X_T^*)}}$  
for  the robust  investor, which is known a priori to be larger than $\underline{\Sc}$ $=$  $\sqrt{ \exp\big(\frac{b^2}{\bar\sigma^2} T\big) -1}$.  
Notice that the optimal stra\-tegy of the robust investor corresponds to the optimal strategy of a simple investor with misspecified volatility $\bar\sigma$.

\begin{figure}
    \centering
    \includegraphics[height=10cm,width=15cm]{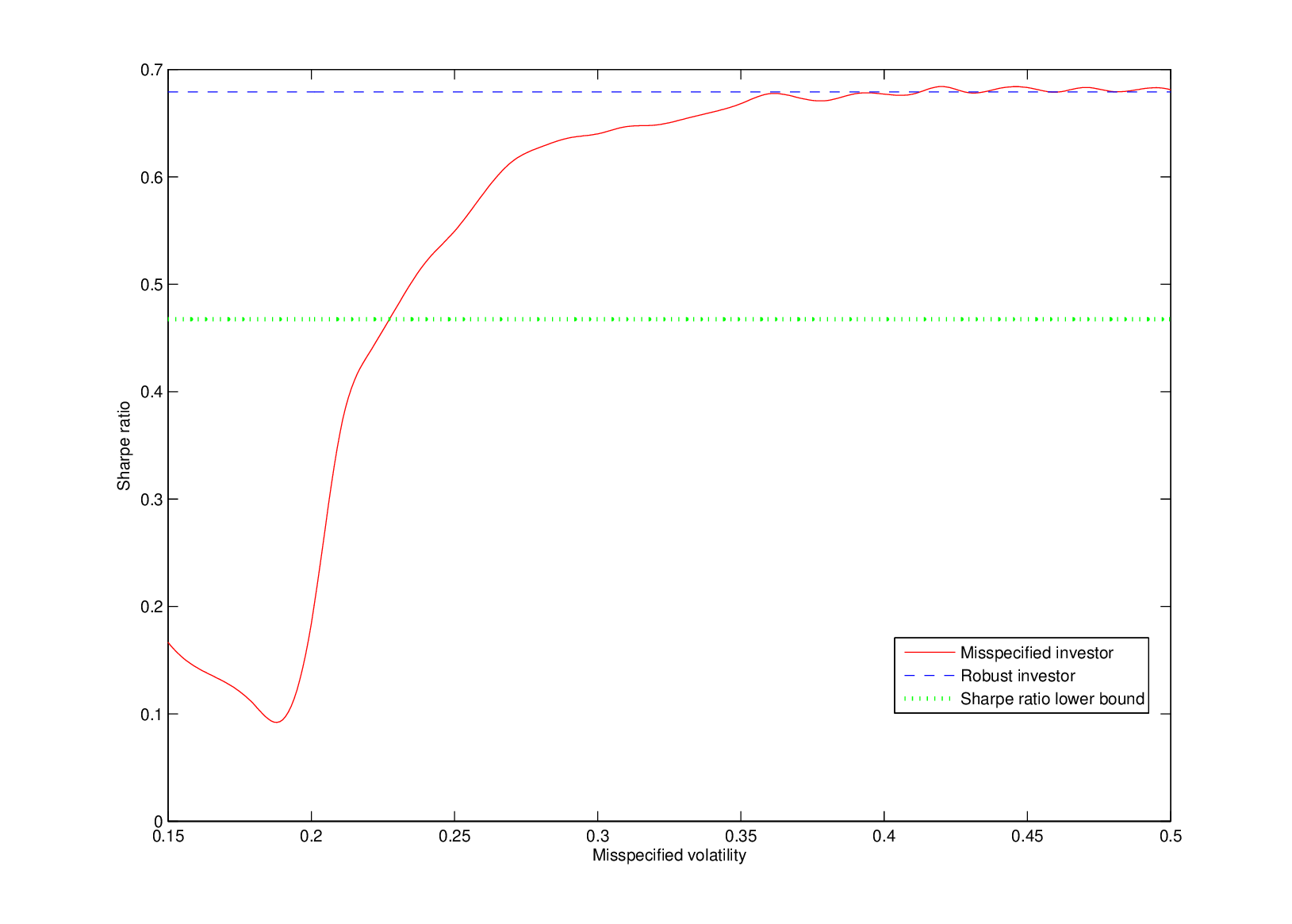}
    \caption{Sharpe ratio $\Sc(\tilde{\alpha})$ for different values of $\tilde\sigma_0$ with parameter values as in Table \ref{tableparam}. }
		\label{fig_vol_sto} 
\end{figure}

\begin{figure}
    \centering
    \includegraphics[height=10cm,width=15cm]{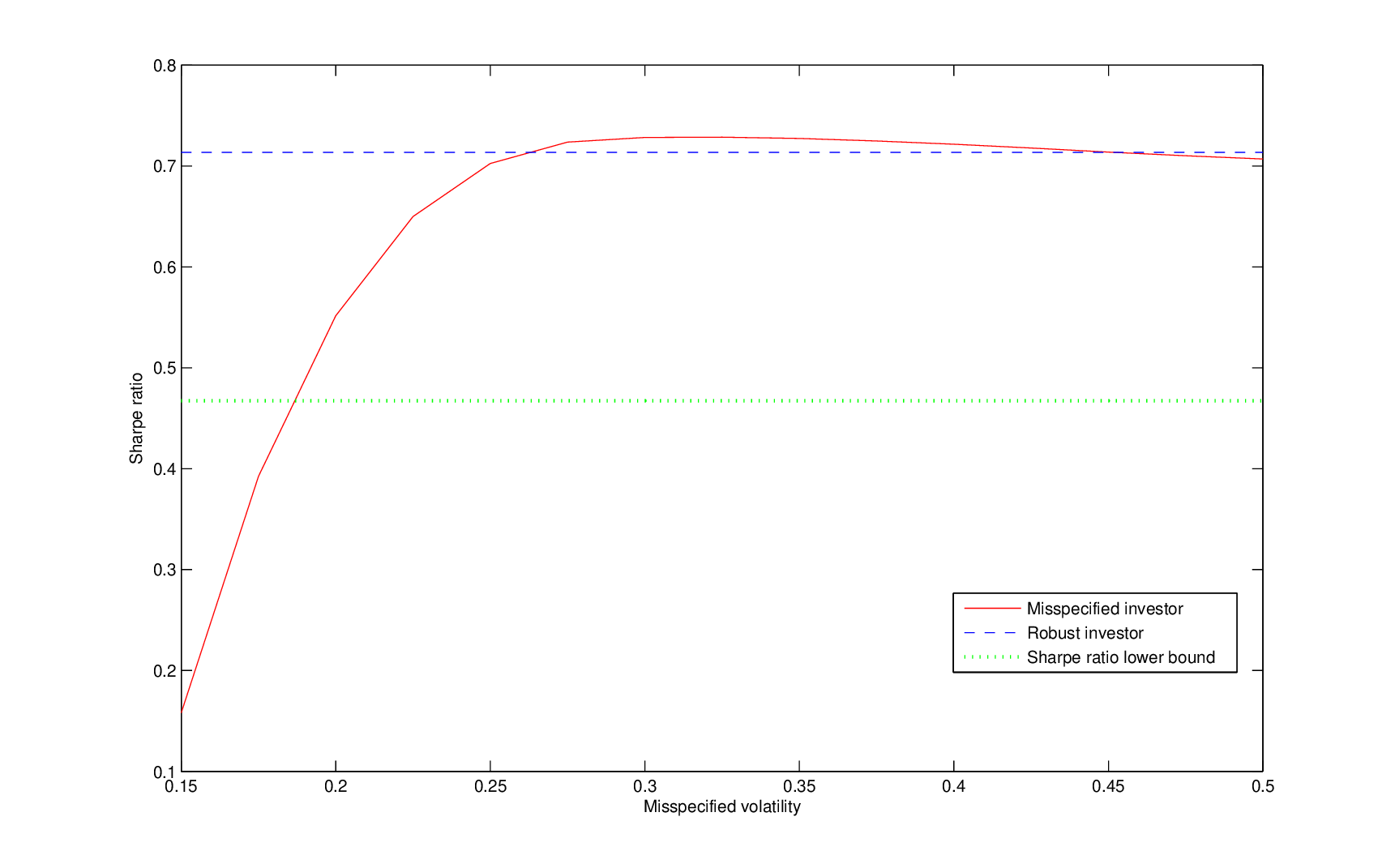}
    \caption{Sharpe ratio $\Sc(\tilde{\alpha})$ for different values of $\tilde\sigma_0$ with parameter values as in Table \ref{tableparam} but with $\kappa$ $=$ $5$, 
    $\eta$ $=$ $0.25$.}
		\label{fig_vol_sto2} 
\end{figure}

Table \ref{tableres}  and Figure \ref{fig_vol_sto}  show the Sharpe ratios of the robust investor and of the simple investor when varying the misspecified volatility $\tilde\sigma_0$.  Since the Sharpe ratios are computed by Monte-Carlo simulations, we also put in Table \ref{tableres} a confidence interval. 
We see that the Sharpe ratio of the robust investor can perform noticeably better than the one of the simple investor who uses a misspecified volatility: this gap is all the more important as  the misspecified volatility  is far from the stationary value $\sigma_\infty$ of the true volatility, for example when $\tilde\sigma_0$ $=$ $\underline{\sigma}$.  On the other hand,  we notice that the Sharpe ratio of the simple  investor is obviously equal to the one of the robust investor when the misspecified volatility $\tilde\sigma_0$ is equal to the worst-case  scenario of volatility $\bar\sigma$. 
Let us mention  that the outperformance of the robust strategies with respect to the misspecified Black-Scholes strategies is illustrated in our  example for a specific choice of parameters. However, it may happen that when the vol-of-vol $\eta$ is low, and/or the speed of mean-reversion $\kappa$ is high, then the Black-Scholes investor using a misspecified volatility closed to the long-run volatility $\sigma_\infty$ will perform better than the robust investor, 
as shown through Table \ref{tableres2} and Figure \ref{fig_vol_sto2}, where we have used $\eta=0.25$ and $\kappa=5$ while keeping the other parameters as specified in Table \ref{tableparam}.


\begin{table}
\begin{center} 
\begin{tabular}{|c|c|c|c|c|c|c|c|}
\hline
$b$ & $\kappa$ & $\eta$ & $\sigma_0$ & $\underline{\sigma}$ & $\sigma_\infty$ & $\bar\sigma$ & $\rho$   \\
\hline
20$\%$  &  2 & 1& 30$\%$ & 15$\%$  & 30$\%$ & 45$\%$  &  -0.7   \\
\hline
\end{tabular} 

\vspace{2mm}

\caption{Parameter values used in the bounded Heston stochastic volatility model. }
\label{tableparam}
\end{center} 
\end{table}




 \begin{table}
\begin{center}
\begin{adjustbox}{max width=\textwidth} 
\begin{tabular}{|c||c|c|c|c|c|}
\hline
$\tilde\sigma_0$   & $\underline{\sigma}$ & 20$\%$ & $\sigma_{\infty}$ & $\bar\sigma$ & 50$\%$\\ 
\hline
$\underline{\Sc}$ & 0.4673  & 0.4673 & 0.4673 & 0.4673 & 0.4673  \\
\hline
$\Sc(\alpha^*)$ & 0.6831 & 0.6831 & 0.6831 & 0.6831 & 0.6831  \\
\hline
95\% confidence interval for $\Sc(\alpha^*)$  & [0.6817,0.6844] & [0.6817,0.6844] & [0.6817,0.6844] & [[0.6817,0.6844] & [0.6817,0.6844] \\
\hline
$\Sc(\tilde{\alpha})$ & 0.1666 & 0.1839 & 0.64 & 0.6831  & 0.6809   \\
\hline
95\% confidence interval for $\Sc(\tilde{\alpha})$  & [0.1662,0.1669] & [0.1835,0.1842] & [0.6387,0.6412] & [0.6817,0.6844] & [0.6795,0.6822]  \\
\hline
\end{tabular}
\end{adjustbox} 
\vspace{2mm}
\caption{\small}{Sharpe ratios $\Sc(\alpha^*)$  of the robust investor and  $\Sc(\tilde{\alpha})$ of the  investor for different misspecified values of $\tilde\sigma_0$ and parameter values as in Table \ref{tableparam}. }
\label{tableres}
\end{center} 
\end{table}

\begin{table}
\begin{center}
\begin{adjustbox}{max width=\textwidth} 
\begin{tabular}{|c||c|c|c|c|c|}
\hline
$\tilde\sigma_0$   & $\underline{\sigma}$ & 20$\%$ & $\sigma_{\infty}$ & $\bar\sigma$ & 50$\%$\\ 
\hline
$\underline{\Sc}$ & 0.4673  & 0.4673 & 0.4673 & 0.4673 & 0.4673  \\
\hline
$\Sc(\alpha^*)$ & 0.7135 & 0.7135 & 0.7135 & 0.7135 & 0.7135  \\
\hline
95\% confidence interval for $\Sc(\alpha^*)$  & [0.7108,0.7136] & [0.7108,0.7136] & [0.7108,0.7136] & [0.7108,0.7136] & [0.7108,0.7136]\\
\hline
$\Sc(\tilde{\alpha})$ & 0.1581 & 0.5515 & 0.7282 & 0.7135  & 0.7069   \\
\hline
95\% confidence interval for $\Sc(\tilde{\alpha})$  & [0.1578,0.1584] & [0.5503,0.5525] & [0.7273,0.7301] & [0.7108,0.7136] & [0.7075,0.7102]  \\
\hline
\end{tabular}
\end{adjustbox} 
\vspace{2mm}
\caption{\small}{Sharpe ratios $\Sc(\alpha^*)$  of the robust investor and  $\Sc(\tilde{\alpha})$ of the  investor for different misspecified values of $\tilde\sigma_0$ and parameter values as in Table \ref{tableparam} but with $\kappa$ $=$ $5$, $\eta$ $=$ $0.25$.}
\label{tableres2}
\end{center} 
\end{table}

 \subsection{A stochastic correlation model}

We consider a market with two risky assets, and motivated by the model in \cite{chiwon14}, assume that the true dynamics of the stock price $S$ $=$ 
$(S^1,S^2)$ is governed by 
\beq
dS_t &=& {\rm diag}(S_t) \big[bdt + \varsigma(\varrho_t) dW_t \big] \nonumber \\
&=& \left( 
\begin{array}{l} 
S_t^1 \big[ b_1 dt + \sigma_1 \sqrt{1 - \varrho_t^2} dW_t^1 + \sigma_1 \varrho_t dW_t^2\big] \\
S_t^2 \big[ b_2 dt + \sigma_2 dW_t^2\big]
\end{array}
\right), \label{stocorellS}
\enq
where $b$ $=$ $(b_1,b_2)$,   $\sigma_1>0$, $\sigma_2$ $>$ $0$ are known constants, and $(\varrho_t)$ is a stochastic correlation process valued in 
$[0,\bar\varrho]$, with a known positive constant $\bar\varrho$ $<$ $1$, and governed by a Wright-Fisher dynamics
\beq \label{stocorellrho}
d\varrho_t &=& \kappa (\varrho_\infty - \varrho_t) dt +  \eta \sqrt{\varrho_t(\bar\varrho - \varrho_t)} d\tilde W_t,
\enq
where  $\kappa$ $\geq$ $0$, $\varrho_\infty$ $\in$ $[0,\bar\varrho]$,  $\eta$ $>$ $0$, and $\tilde W$ is a Brownian motion, assumed here for simplicity, to be  independent of the two dimensional Brownian motion $W$ $=$ $(W^1,W^2)$ under the real probability measure $\P$.    
 
We now consider a simple investor who knows the drifts $b_i$, the volatilities  $\sigma_i$, 
hence the corresponding instantaneous Sharpe ratios $\beta_i$ $=$ $b_i/\sigma_i$, of the two assets $i$ $=$ $1,2$, but specifies  incorrectly the correlation by  considering that it is equal to a constant $\tilde\varrho_0$ $\in$ $(-1,1)$.  Therefore, from the result in \cite{zholi00} or as a particular case of our paragraph \ref{seccorrel}  when $\Theta$ is reduced to the singleton 
$\tilde\varrho_0$,   the optimal mean-variance portfolio strategy of this   ``misspecified" investor with risk-aversion parameter $\lambda$ $>$ $0$, and initial capital $x_0$ is given by:
\beqs
\tilde\alpha_t &=& \Big[ x_0 + \frac{1}{2\lambda} \exp\big(   \tilde R_0 T  \big) - \tilde X_t \Big] \tilde\Sigma_0^{-1} b, \;\;\; 0 \leq t \leq T,
\enqs
where 
\beqs
\tilde\Sigma_0 \;:= \; \gamma(\tilde\varrho_0) \;  = \; 
\Big( 
\begin{array}{cc}
\sigma_1^2   & \sigma_1\sigma_2 \tilde\varrho_0 \\
\sigma_1\sigma_2 \tilde\varrho_0 & \sigma_2^2 
\end{array}
\Big), & & 
\tilde\Sigma_0^{-1}b \;  = \; \frac{1}{1-\tilde\varrho_0^2}
\left( 
\begin{array}{c}
\frac{\beta_1 - \beta_2\tilde\varrho_0}{\sigma_1}     \\
\frac{\beta_2 -  \beta_1\tilde\varrho_0}{\sigma_2}  
\end{array}
\right)
\enqs
\beqs
\tilde R_0  & := & b\trans\tilde\Sigma_0^{-1}b \; = \; \frac{1}{1-\tilde\varrho_0^2} \big( \beta_1^2 + \beta_2^2 - 2 \beta_1 \beta_2 \tilde\varrho_0 \big), 
\enqs 
and $\tilde X$  is the wealth process with  feedback strategy $\tilde\alpha$, governed under the real probability measure $\P$ by
\beqs
d\tilde X_t &=& \tilde\alpha_t\trans b dt + \tilde\alpha_t\trans\varsigma(\varrho_t) dW_t.  
\enqs
Its expected return under $\P$ is then governed by
\beqs
d \E[\tilde X_t] &=& b\trans\E[\tilde\alpha_t] dt \; = \; \tilde R_0  \Big[ x_0 + \frac{1}{2\lambda} \exp\big(   \tilde R_0 T  \big) - \E[\tilde X_t] \Big] dt, 
\enqs
which gives the excess of expected return at $T$:
\beqs
\E[\tilde X_T ] - x_0 &=& \frac{1}{2\lambda} \big[  \exp\big(   \tilde R_0 T  \big) - 1 \big]. 
\enqs
The variance risk  of $\tilde X_T$ under $\P$ is not explicit, but can be approximated by $N$ Monte-Carlo simulations $(\tilde X^i)_{i=1,\ldots,N}$  of 
$\tilde X$ under $\P$ via:
\beqs
{\rm Var}(\tilde X_T) & \simeq & \frac{1}{N} \sum_{i=1}^N \big( \tilde X_T^i - \E[\tilde X_T] \big)^2.
\enqs  
We  can then compute the  Sharpe ratio $\Sc(\tilde\alpha)$ $=$ $\frac{ \E[\tilde X_T] - x_0}{\sqrt{ {\rm Var}(\tilde X_T)}}$  for  the ``misspecified" investor. 

The model parameters used in the simulations of $\tilde X$ in the stochastic correlation model \reff{stocorellS}-\reff{stocorellrho}  are given in Table \ref{tableparam2}. 
Again, we used the simulation method as in \cite{deedel98} for dealing with the discretization of the Wright-Fisher process for the correlation, which means that when a correlation trajectory breachs the bounds $0$ or $\bar\varrho$, we project its value according to its closest neighbor on $[0,\bar\varrho]$. 
We fix an investment horizon $T$ $=$ $1$ year, a risk-aversion parameter $\lambda$ $=$ $5$, and use $N$ $=$ $500 000$ simulations  for each set of parameters.

On the other hand, let us consider a robust investor with risk-aversion parameter $\lambda$,  initial capital $x_0$.  By taking the parameters in Table \ref{tableparam2}, we notice that $\varrho_0^+$ $=$ $\beta_2/\beta_1$ $\in$ $[0,\bar\varrho]$, and thus from the result in Theorem \ref{procor}, 
her/his robust efficient portfolio strategy $\alpha^*$ $=$ $\alpha^{*,\lambda}$ is given by
\beqs
\alpha_t^* &=& 
\left(
\begin{array}{c}
\Big[ x_0 + \frac{1}{2\lambda} \exp\big(  \beta_1^2  T  \big) - X_t^* \Big] \frac{b_1}{\sigma_1^2}  \\
0
\end{array}
\right),  \;\; 0 \leq t \leq T,
\enqs
and her/his wealth process $X^*$ is governed under the real probability measure $\P$ by
\beq
d X_t^* &=& (\alpha_t^*)\trans b dt + (\alpha_t^*)\trans\varsigma(\varrho_t) dW_t \nonumber \\
&=&  \Big[ x_0 + \frac{1}{2\lambda} \exp\big(  \beta_1^2  T  \big) - X_t^* \Big] \beta_1^2 dt + 
\Big[ x_0 + \frac{1}{2\lambda} \exp\big(  \beta_1^2  T  \big) - X_t^* \Big] \beta_1 \sqrt{1-\rho_t^2} dW_t^1 \nonumber \\
& & \;\;\; + \;  \Big[ x_0 + \frac{1}{2\lambda} \exp\big(  \beta_1^2  T  \big) - X_t^* \Big] \beta_1 \rho_t dW_t^2.  \label{X*}
\enq
The excess of expected return  under  $\P$  is explicitly given by 
\beq \label{excessrho}
\E[X_t^*] - x_0 &=& \frac{1}{2\lambda} \big[  \exp\big(  \beta_1^2 T  \big) - \exp\big(  \beta_1^2 (T-t)  \big) \big], \;\;\; 0 \leq t \leq T. 
\enq
and we can actually  compute explicitly in this case the variance risk  of $X_t^*$ under the real probability measure.  Indeed, denoting by 
$Y_t^*$ $=$ $X_t^*-\E[X_t^*]$, we see from \reff{X*}-\reff{excessrho} that 
\beqs
dY_t^* &=& -\beta_1^2 Y_t^* dt + \big(  \frac{1}{2\lambda} e^{\beta_1^2(T-t)} - Y_t^*\big) 
\big[ \beta_1 \sqrt{1-\rho_t^2} dW_t^1  + \beta_1 \rho_t dW_t^2 \big],
\enqs
so that by It\^o's formula, and taking expectation under $\P$: 
\beqs
d \E|Y_t^*|^2 &=& \big( -  \beta_1^2 \E|Y_t^*|^2 + \frac{\beta_1^2}{4\lambda^2} e^{2\beta_1^2 (T-t)} \big) dt.  
\enqs
It follows that
\beqs \label{varrho}
{\rm Var}(X_t^*) &=& \E|Y_t^*|^2 \; = \; \frac{e^{2\beta_1^2(T-t)}}{4\lambda^2} \big( e^{\beta_1^2 t} - 1 \big), \;\;\; 0 \leq t \leq T. 
\enqs
In particular, we deduce  the Sharpe ratio of the robust investor:
\beqs
\Sc(\alpha^*) &=& \frac{ \E[X_T^*] - x_0}{\sqrt{ {\rm Var}(X_T^*)}} \; = \; \sqrt{ \exp\big(\beta_1^2 T\big) -1} \; = \; \underline{\Sc},
\enqs
which means that in the case when $\varrho_0^+$ $\in$ $[0,\bar\varrho]$, the Sharpe ratio attains its lower bound $\underline{\Sc}$.  
Notice that the optimal strategy of the robust investor is equal  to the optimal strategy of a simple investor with misspecified correlation 
$\tilde\varrho_0$ $=$ $\varrho_0^+$.

\begin{figure}
    \centering
    \includegraphics[height=6.4cm,width=15cm]{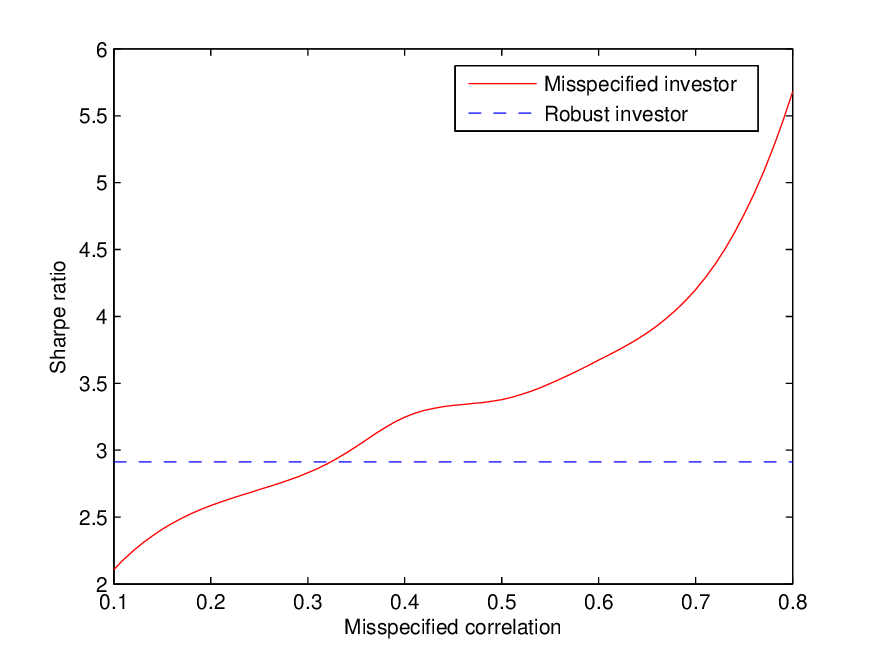}
    \caption{Sharpe ratio  $\Sc(\tilde{\alpha})$  for different values of $\tilde\varrho_0$}   
		\label{fig_correl_sto1} 
\end{figure}

Table \ref{tablerescor}  and Figure \ref{fig_correl_sto1} show the Sharpe ratios of the robust investor and of the simple investor when varying the misspecified correlation $\tilde\varrho_0$ (since the Sharpe ratio of the simple investor  is computed by Monte-Carlo simulations, we also put in Table \ref{tablerescor} its confidence interval at level $95\%$). 
They obviously coincide by definition when the misspecified correlation $\tilde\varrho_0$ is equal to $\varrho_0^+$ (here equal to $\beta_2/\beta_1$ 
$=$ $1/3$). On the  other hand,  we see that the Sharpe ratio of the robust investor may perform worse than the one of the simple investor, especially when the misspecified correlation  
$\tilde\varrho_0$ is close from the true stationary correlation $\varrho_\infty$ (and when the vol-of-correl $\eta$ is low, and/or the speed of mean-reversion $\kappa$ is high),  but  performs better when  $\tilde\varrho_0$  is smaller than  $\varrho_0^+$.


\begin{table}
\begin{center} 
\begin{tabular}{|c|c|c|c|c|c|c|}
\hline
$\beta_1$ & $\beta_2$ &  $\varrho_0$ &  $\bar\varrho$ & $\kappa$ & $\varrho_\infty$ & $\eta$  \\
\hline
1.5 &  0.5  & 0.7  &  0.95   &   5  &  0.7  & 20$\%$     \\
\hline
\end{tabular} 

\vspace{2mm}

\caption{Parameter values used in the stochastic correlation model}
\label{tableparam2}
\end{center} 
\end{table} 




\begin{table}
\begin{center}
\begin{adjustbox}{max width=\textwidth}  
\begin{tabular}{|c||c|c|c|c|}
\hline
$\tilde\varrho_0$  & 0.1  & $\varrho_0^+$ $=$ $1/3$ & $\varrho_{\infty}$ & 0.8   \\ 
\hline
$\Sc(\alpha^*)$ $=$ $\underline{\Sc}$   & 2.9134   & 2.9134   &  2.9134   &  2.9134   \\
\hline
$\Sc(\tilde{\alpha})$ & 2.1085 & 2.9134   &  4.2008    & 5.6798    \\
\hline
95\% confidence interval for $\Sc(\tilde{\alpha})$  & [2.1043,2.1126] & [2.9076,2.9191]   &  [4.1925,4.2090]    & [5.6686,5.6909]  \\
\hline
\end{tabular}
\end{adjustbox}

\vspace{2mm}

\caption{\small}{Sharpe ratios $\Sc(\alpha^*)$  of the robust investor and  $\Sc(\tilde{\alpha})$ of the  investor for different misspecified values of $\tilde{\varrho}_{0}$. }
\label{tablerescor}
\end{center} 
\end{table}  


 \clearpage

\appendix





\section{Appendix: Differentiability on Wasserstein space and It\^o's formula}

\setcounter{equation}{0} \setcounter{Assumption}{0}
\setcounter{Theorem}{0} \setcounter{Proposition}{0}
\setcounter{Corollary}{0} \setcounter{Lemma}{0}
\setcounter{Definition}{0} \setcounter{Remark}{0}

We first recall  the notion of derivative with respect to a probability measure, as introduced by P.L. Lions in his course at Coll\`ege de France, and detailed   in the lecture notes  \cite{car12}.

This notion is based on the lifting of functions $u$ $:$ $\Pc_{_2}(\R)$ $\rightarrow$ $\R$ into functions $U$ defined on $L^2(\Gc;\R)$ $=$ 
$L^2(\Omega,\Gc,\P;\R)$ (the set of square-integrable random variables on some probability space $(\Omega,\Gc,\P)$) 
by $U(X)$ $=$ $u(\Lc(X))$, where $\Lc(X)$ is the law of $X$ on $(\Omega,\Gc,\P)$. We say that $u$ is differentiable (resp. $\Cc^1$) on
$\Pc_{_2}(\R)$ if the lift $U$ is Fr\'echet differentiable (resp. Fr\'echet differentiable with continuous derivatives) on  $L^2(\Gc;\R)$. In this case, the 
Fr\'echet derivative $[DU](X)$, which is identified  as an element $DU(X)$ of $L^2(\Gc;\R)$  by Riesz' theorem through the relation: $[DU](X)(Y)$ $=$ $\E[DU(X)Y]$,  can be represented as
\beq \label{Uu1}
DU(X) &=& \partial_\mu u(\Lc(X))(X),
\enq
for some function  $\partial_\mu u(\Lc(X))$ $:$ $\R$ $\rightarrow$ $\R$,  which is  called derivative of $u$ at $\mu$ $=$ $\Lc(X)$.  
Moreover, $\partial_\mu u(\mu)$ $\in$ $L^2(\mu)$ for $\mu$ $\in$ $\Pc_{_2}(\R)$ $=$ $\{\Lc(X), X \in L^2(\Gc;\R)\}$. 
We say that $u$ is partially $\Cc^2$ if it is $\Cc^1$, and one can find, for any $\mu$ $\in$ $\Pc_{_2}(\R)$, 
a continuous version of the mapping $x\in\R$ $\mapsto$ $\partial_\mu u(\mu)(x)$, such that the mapping
$(\mu,x)$ $\in$ $\Pc_{_2}(\R)\times\R$ $\mapsto$ $\partial_\mu u(\mu)(x)$  is continuous at any point $(\mu,x)$ such that $x$ $\in$ Supp$(\mu)$, and if for any $\mu$ $\in$ $\Pc_{_2}(\R)$, the mapping
$x$ $\in$ $\R$ $\mapsto$  $\partial_\mu u(\mu)(x)$ is differentiable, its derivative being jointly continuous at any point  $(\mu,x)$ such that $x$ $\in$ Supp$(\mu)$. The gradient is then denoted by $\partial_x  \partial_\mu u(\mu)(x)$.  

For example, consider a  linear function: $u(\mu)$ $=$ $\int \varphi(x) \mu(dx)$. Its lifted function is $U(X)$ $=$ $\E[\varphi(X)]$, whose Fr\'echet derivative is given by: $[DU](X)(Y)$ $=$ $\E[D_x\varphi(X).Y]$, from which we see that $\partial_\mu u(\mu)$ $=$ $D_x\varphi$, and thus 
$\partial_x\partial_\mu u(\mu)$ $=$ $D^2_{x}\varphi$.  In particular, when $\varphi(x)$ $=$ $x$, i.e., $u(\mu)$ $=$ $\bar\mu$ $:=$ $\int x \mu(dx)$, 
then $\partial_\mu u(\mu)$ 
$=$ $1$. Another example used in this paper is a function $u(\mu)$ $=$ ${\rm Var}(\mu)$ $:=$ $\int (x - \bar\mu)^2 \mu(dx)$. 
In this case, its lifted function is  $U(X)$ $=$ ${\rm Var}(X)$, from which we  see that $DU(X)$ $=$ $2(X- \E[X])$, and thus 
$\partial_\mu u(\mu)(x)$ $=$ $2(x-\bar\mu)$, $\partial_x\partial_\mu u(\mu)(x)$ $=$ $2$.

We next  recall a chain rule (or It\^o's formula)  for functions defined on $\Pc_{_2}(\R)$, proved independently in \cite{buetal14} and \cite{chacridel15}. 
Let us consider a real-valued It\^o process
\beqs
dX_t &=& b_t dt + \sigma_t dW_t, \;\;\; X_0 \in L^2(\Fc_0;\R),
\enqs
where $(b_t)$ and $(\sigma_t)$ are progressively measurable processes with respect to the filtration generated by the $d$-dimensional Brownian motion $W$,  valued respectively  in 
$\R$ and $\R^{1\times d}$, and satisfying the integrability condition:
$\E \Big[ \int_0^T |b_t|^2 + |\sigma_t|^2 dt \Big] $ $<$ $\infty$. 
Let $u$ be a partially $\Cc^2$ function on  $\Pc_{_2}(\R)$.  Then, for all $t$ $\in$ $[0,T]$,
\beq 
u(\Lc(X_t)) &=& u(\Lc(X_0))  \; + \; \int_0^t  \E \big[ \partial_\mu u(\Lc(X_s))(X_s) b_s     \nonumber \\
& & \;\;\;\;\;\;\; \hspace{2cm}  + \;  \frac{1}{2}   \partial_x\partial_\mu u(\Lc(X_s))(X_s) |\sigma_s|^2  \big]  ds.  \label{Ito}
\enq


\small

\bibliographystyle{plain}

\end{document}